\documentstyle[prd,preprint,tighten,aps,eqsecnum,%
amssymb,newlfont,epsfig]{revtex}
\begin{document}
\draft
\preprint{
\begin{tabular}{r}
UWThPh-1997-18\\
DFTT 54/97\\
IASSNS-AST 97/52\\
hep-ph/9710209
\end{tabular}
}
\title{BOUNDS ON LONG-BASELINE
$\bar\nu_e\to\bar\nu_e$ AND 
$
\stackrel{\makebox[0pt][l]
{$\hskip-3pt\scriptscriptstyle(-)$}}{\nu_{\mu}}
\to\stackrel{\makebox[0pt][l]
{$\hskip-3pt\scriptscriptstyle(-)$}}{\nu_{e}}
$
TRANSITION PROBABILITIES}
\author{S.M. Bilenky}
\address{Joint Institute for Nuclear Research, Dubna, Russia, and\\
Institute for Advanced Study, Princeton, N.J. 08540}
\author{C. Giunti}
\address{INFN, Sezione di Torino, and Dipartimento di Fisica Teorica,
Universit\`a di Torino,\\
Via P. Giuria 1, I--10125 Torino, Italy}
\author{W. Grimus}
\address{Institute for Theoretical Physics, University of Vienna,\\
Boltzmanngasse 5, A--1090 Vienna, Austria}
\maketitle
\begin{abstract}
We discuss long-baseline neutrino oscillations in the framework of the 
two 4-neutrino schemes which can accommodate all
existing neutrino oscillation data.
Negative results of short-baseline reactor and accelerator experiments
allow to obtain rather strong bounds on the long-baseline $\bar\nu_e
\to \bar\nu_e$ and 
$
\stackrel{\makebox[0pt][l]
{$\hskip-3pt\scriptscriptstyle(-)$}}{\nu_{\mu}}
\to\stackrel{\makebox[0pt][l]
{$\hskip-3pt\scriptscriptstyle(-)$}}{\nu_{e}}
$ 
transition probabilities. 
We consider in detail matter effects and show that the
vacuum bounds are not substantially modified.
We also comment on corresponding bounds in 3-neutrino scenarios.
\end{abstract}


\narrowtext

\section{Introduction}
\label{Introduction}

The problem of neutrino masses and mixing
(see, for example, Refs. \cite{BP78,BP87,Mohapatra-Pal,CWKim})
is the central issue of modern neutrino physics.
A new stage in the investigation of this problem is represented
by long-baseline (LBL) neutrino oscillation experiments:
CHOOZ \cite{CHOOZ},
Palo Verde \cite{PaloVerde},
Kam-Land \cite{Kam-Land},
K2K \cite{K2K} (KEK--Super-Kamiokande),
MINOS \cite{MINOS} (Fermilab--Soudan),
ICARUS \cite{ICARUS} (CERN--Gran Sasso).
The major goal of these experiments is to reach  
the sensitivity of the
``atmospheric neutrino range''
$ 10^{-3} \div 10^{-2} \, \mathrm{eV}^2 $
for the neutrino mass-squared difference.

Concerning reactor experiments, the first LBL experiment CHOOZ
is taking data now,
the Palo Verde LBL experiment
will start later this year
and the Kam-Land experiment is scheduled to start
in the year 2000.
The accelerator LBL experiment K2K
is planned to begin taking data in the year 1999,
whereas
the MINOS and ICARUS experiments
will start in the first years of the next millennium.

What implications for future LBL experiments
can be inferred from the results of short-baseline
(SBL) neutrino oscillation experiments
and solar and atmospheric neutrino experiments?
We will consider here this question in the framework of
two models with four massive neutrinos that
can accommodate all the existing data
on neutrino oscillations.

The results of many neutrino oscillation
experiments are presently available.
Indications in
favour of neutrino oscillations were found
in solar neutrino experiments
(Homestake \cite{Homestake},
Kamiokande \cite{Kam-sol},
GALLEX \cite{GALLEX},
SAGE \cite{SAGE}
and
Super-Kamiokande \cite{SK-sol}),
in atmospheric neutrino experiments
(Kamiokande
\cite{Kam-atm},
IMB
\cite{IMB},
Soudan
\cite{Soudan}
and
Super-Kamiokande \cite{SK-atm})
and
in the LSND experiment \cite{LSND}.
The data 
of these experiments
can be explained
if there is neutrino mixing with the following values of 
neutrino mass-squared differences:
\begin{equation}
\Delta{m}^2_{\mathrm{sun}} \sim 10^{-10} \quad 
\mbox{or} \quad 10^{-5} \, \mathrm{eV}^2 \, ,
\quad
\Delta{m}^2_{\mathrm{atm}} \sim 10^{-3} \div 10^{-2} \,
\mathrm{eV}^2 \, ,
\quad
\Delta{m}^2_{\mathrm{LSND}} \sim 1 \, \mathrm{eV}^2 \,
.
\label{001}
\end{equation}
The two estimates of $\Delta{m}^2_{\mathrm{sun}}$ refer to the
vacuum oscillation solution \cite{barger}
and the MSW solution \cite{MSW,SOLMSW}, respectively,
of the solar neutrino deficit. The estimate of
$\Delta{m}^2_{\mathrm{atm}}$ derives from the zenith angle
variation of the atmospheric neutrino anomaly. It has so far
only been observed by Kamiokande \cite{Kam-atm}
and Super-Kamiokande \cite{SK-atm}.
From the analysis of the data of the LSND experiment
and the negative results of other SBL experiments
(the strongest limits are provided by
the Bugey \cite{Bugey95} and BNL E776 \cite{BNLE776}
experiments),
it follows that
\begin{equation}\label{range}
0.3
\lesssim
\Delta{m}^2_{\mathrm{LSND}}
\lesssim
2.2 \: \mathrm{eV}^2 \, .
\end{equation}

There are also data of many different reactor
and accelerator short-baseline neutrino oscillation
experiments in which no indication in favour of oscillations
was found
(see the reviews in Ref. \cite{Boehm-Vannucci}).

Three different scales of mass-squared differences
require schemes with (at least) four massive neutrinos
\cite{four,BGKP,BGG96,OY96}
(see, however,
Refs. \cite{AP97,CF97-MR97,FLMS97}
for scenarios with three massive neutrinos and Ref. \cite{baksan}
for comments on these scenarios).
In Refs. \cite{BGG96,OY96} all possible 4-neutrino mass spectra
with the solar, atmospheric and LSND mass-squared difference scales 
were considered.
It was shown that only two of these schemes
are compatible with all the existing data.
In these two schemes
the four neutrino masses
are divided into two pairs of close masses
separated by a gap of the order of 1 eV,
which gives
$ \Delta{m}^2_{\mathrm{LSND}} = \Delta{m}^2_{41} \sim 1 \mathrm{eV}^2 $:
\begin{equation}\label{AB}
\mbox{(A)}
\qquad
\underbrace{
\overbrace{m_1 < m_2}^{\mathrm{atm}}
\ll
\overbrace{m_3 < m_4}^{\mathrm{sun}}
}_{\mathrm{LSND}}
\qquad \mbox{and} \qquad
\mbox{(B)}
\qquad
\underbrace{
\overbrace{m_1 < m_2}^{\mathrm{sun}}
\ll
\overbrace{m_3 < m_4}^{\mathrm{atm}}
}_{\mathrm{LSND}}
\;.
\end{equation}
In scheme A,
$\Delta{m}^{2}_{21} \equiv \Delta m^2_{\mathrm{atm}}$
is relevant
for the explanation of the atmospheric neutrino anomaly
and
$\Delta{m}^{2}_{43} \equiv \Delta m^2_{\mathrm{sun}}$
is relevant
for the suppression of solar $\nu_e$'s.
In scheme B,
the r\^{o}les of
$\Delta{m}^{2}_{21}$
and
$\Delta{m}^{2}_{43}$
are reversed.

In the framework of the schemes (\ref{AB}),
the probabilities of SBL transitions 
have the form \cite{BGG96}
\begin{eqnarray}
&&
\label{Pab}
P^{(\mathrm{SBL})}_{\stackrel{\makebox[0pt][l]
{$\hskip-3pt\scriptscriptstyle(-)$}}{\nu_{\alpha}}
\to\stackrel{\makebox[0pt][l]
{$\hskip-3pt\scriptscriptstyle(-)$}}{\nu_{\beta}}}
=
\frac{1}{2}
\,
A_{\alpha;\beta}
\left( 1 - \cos\frac{ \Delta{m}^{2} \, L }{ 2 \, p } \right)
\qquad
(\beta\neq\alpha)
\,,
\\
&&
\label{Paa}
P^{(\mathrm{SBL})}_{\stackrel{\makebox[0pt][l]
{$\hskip-3pt\scriptscriptstyle(-)$}}{\nu_{\alpha}}
\to\stackrel{\makebox[0pt][l]
{$\hskip-3pt\scriptscriptstyle(-)$}}{\nu_{\alpha}}}
=
1
-
\frac{1}{2}
\,
B_{\alpha;\alpha}
\left( 1 - \cos\frac{ \Delta{m}^{2} \, L }{ 2 \, p } \right)
\,,
\end{eqnarray}
which are similar to the standard two-neutrino transition probabilities.
From now on we use the notation
$ \Delta{m}^{2} \equiv \Delta{m}^{2}_{41} \equiv m^2_4 - m^2_1 $
for the SBL mass-squared difference,
$L$ is the source--detector distance,
$p$ is the neutrino momentum
and the oscillation amplitudes are given by
\begin{eqnarray}
A_{\alpha;\beta}
& = &
4 \left| \sum_{k=1,2} U_{{\beta}k} U_{{\alpha}k}^{*} \right|^2
=
4 \left| \sum_{k=3,4} U_{{\beta}k} U_{{\alpha}k}^{*} \right|^2
\,,
\label{Aab}
\\
B_{\alpha;\alpha}
& = &
4
\left( \sum_{k=1,2} |U_{{\alpha}k}|^2 \right)
\left( 1 - \sum_{k=1,2} |U_{{\alpha}k}|^2 \right)
\nonumber
\\
& = &
4
\left( \sum_{k=3,4} |U_{{\alpha}k}|^2 \right)
\left( 1 - \sum_{k=3,4} |U_{{\alpha}k}|^2 \right)
\label{Baa}
\,,
\end{eqnarray}
where $U$ is
the unitary mixing matrix that connects flavour and sterile fields
with the fields of neutrinos with definite masses:
\begin{equation}\label{mixing}
\nu_{{\alpha}L}
=
\sum_{k=1}^{4}
U_{{\alpha}k}
\,
\nu_{kL}
\qquad
(\alpha=e,\mu,\tau,s)
\,.
\end{equation}

Eqs.(\ref{Paa}) and (\ref{Baa}) and SBL disappearance data lead
to further information on the schemes A and B.
From the exclusion plots obtained in
the Bugey \cite{Bugey95},
CDHS \cite{CDHS84} and CCFR \cite{CCFR84} disappearance experiments,  
it follows that
\begin{equation} \label{B0}
B_{\alpha;\alpha} \leq B_{\alpha;\alpha}^{0}
\qquad
( \alpha = e , \mu )
\;.
\end{equation}
The values of these upper bounds depend on
$\Delta{m}^2$.
We have considered the range
\begin{equation}
10^{-1} \, \mathrm{eV}^2
\leq \Delta{m}^{2} \leq
10^3 \, \mathrm{eV}^2
\,.
\label{widerange}
\end{equation}
In this range of $\Delta{m}^{2}$ the amplitude $B_{e;e}^{0}$ is small,
whereas
$B_{\mu;\mu}^{0}$
is small for
$ \Delta{m}^{2} \gtrsim 0.3 \, \mathrm{eV}^2 $.

Taking into account 
the results of solar and atmospheric neutrino
experiments,
for the elements of the mixing matrix we have
the following bounds
in the two schemes (\ref{AB}):
\begin{eqnarray}
\mathrm{(A)} \qquad & c_{e} \leq a^{0}_{e} \,, & 
\qquad c_{\mu} \geq 1 - a^{0}_{\mu} \,, \label{A}\\
\mathrm{(B)} \qquad & \!\qquad c_{e} \geq 1 - a^{0}_{e} \,, & 
\qquad c_{\mu} \leq a^{0}_{\mu} \,, \label{B}
\end{eqnarray}
where 
\begin{equation}\label{defca}
c_{\alpha}
\equiv
\sum_{k=1,2} |U_{{\alpha}k}|^2
\end{equation}
and
\begin{equation} \label{a0}
a^{0}_{\alpha} = \frac{1}{2}
\left(1-\sqrt{1-B_{\alpha;\alpha}^{0}}\,\right)
\qquad (\alpha = e,\mu)
\,.
\end{equation}
The values of
$a^{0}_{e}$
and
$a^{0}_{\mu}$
are given in Fig. 1 of Ref. \cite{BBGK}
(one can see that
$ a^{0}_e \lesssim 4 \times 10^{-2} $
for $\Delta{m}^{2}$
in the range (\ref{widerange})
and
$ a^{0}_\mu \lesssim 10^{-1} $
for
$
\Delta{m}^{2} \gtrsim 0.5 \, \mathrm{eV}^2
$).

In the following we will use also the bounds
on the amplitude of
$
\stackrel{\makebox[0pt][l]
{$\hskip-3pt\scriptscriptstyle(-)$}}{\nu_{\mu}}
\to\stackrel{\makebox[0pt][l]
{$\hskip-3pt\scriptscriptstyle(-)$}}{\nu_{e}}
$
transition
which can be obtained from exclusion plots of the
BNL E734
\cite{BNLE734},
BNL E776
\cite{BNLE776}
and
CCFR
\cite{CCFR97}
appearance experiments. Thus,
we can write
\begin{equation} \label{A0}
A_{\mu;e} \leq A_{\mu;e}^{0}
\,,
\end{equation}
where the value of $A_{\mu;e}^{0}$
corresponding to each value of $\Delta m^2$
can be obtained
from the combination of these exclusion plots.

In this paper we will show that,
in the framework of the two schemes (\ref{AB}),
rather strong limits on the
LBL
$ \bar\nu_e \to \bar\nu_e $
and
$
\stackrel{\makebox[0pt][l]
{$\hskip-3pt\scriptscriptstyle(-)$}}{\nu_{\mu}}
\to\stackrel{\makebox[0pt][l]
{$\hskip-3pt\scriptscriptstyle(-)$}}{\nu_{e}}
$
vacuum transition probabilities are obtained. The first of these
channels will be investigated 
in the CHOOZ, Palo Verde and Kam-Land experiments
and the second one
by the K2K, MINOS and ICARUS collaborations.
There are no similar limits
on the probability of
$
\stackrel{\makebox[0pt][l]
{$\hskip-3pt\scriptscriptstyle(-)$}}{\nu_{\mu}}
\to\stackrel{\makebox[0pt][l]
{$\hskip-3pt\scriptscriptstyle(-)$}}{\nu_{\mu}}
$
and
$
\stackrel{\makebox[0pt][l]
{$\hskip-3pt\scriptscriptstyle(-)$}}{\nu_{\mu}}
\to\stackrel{\makebox[0pt][l]
{$\hskip-3pt\scriptscriptstyle(-)$}}{\nu_{\tau}}
$
oscillations.

Furthermore, we will consider in this paper the LBL
transition probabilities of the 
$ \bar\nu_e \to \bar\nu_e $
and
$
\stackrel{\makebox[0pt][l]
{$\hskip-3pt\scriptscriptstyle(-)$}}{\nu_{\mu}}
\to\stackrel{\makebox[0pt][l]
{$\hskip-3pt\scriptscriptstyle(-)$}}{\nu_{e}}
$
channels in the presence of matter.
We will show that the vacuum bounds are not substantially modified 
by matter corrections.

Let us stress that the bounds on the LBL transition
probabilities that we have obtained
are general,
but are heavily based on the existing neutrino oscillation data 
and in particular on the LSND data.
If the LSND indications
in favour of $ \bar\nu_\mu \to \bar\nu_e $
oscillations will not be confirmed
by the future experiments, these bounds will not be valid.

Future measurements by LBL experiments of
$ \bar\nu_e \to \bar\nu_e $
and/or
$
\stackrel{\makebox[0pt][l]
{$\hskip-3pt\scriptscriptstyle(-)$}}{\nu_{\mu}}
\to\stackrel{\makebox[0pt][l]
{$\hskip-3pt\scriptscriptstyle(-)$}}{\nu_{e}}
$
transition probabilities
that violate the bound presented in this paper
would allow to exclude the 4-neutrino schemes (\ref{AB}).

\section{Vacuum Bounds for LBL Neutrino Oscillations}
\label{Vacuum Bounds}

\subsection{The case $\Delta m^2_{\mathrm{sun}} L/2p \ll 1$}
\label{ll}

In scheme A,
the probabilities of
$ \nu_\alpha \to \nu_\beta $
and
$ \bar\nu_\alpha \to \bar\nu_\beta $
transitions
in LBL experiments
are given by
\begin{eqnarray}
&&
P^{(\mathrm{LBL,A})}_{\nu_\alpha\to\nu_\beta}
=
\left|
U_{\beta1}
\,
U_{\alpha1}^{*}
+
U_{\beta2}
\,
U_{\alpha2}^{*}
\,
\exp\!\left(
- i
\frac{ \Delta{m}^{2}_{21} \, L }{ 2 \, p }
\right)
\right|^2
+
\left|
\sum_{k=3,4}
U_{{\beta}k}
\,
U_{{\alpha}k}^{*}
\right|^2
\;,
\label{plba1}
\\
&&
P^{(\mathrm{LBL,A})}_{\bar\nu_\alpha\to\bar\nu_\beta}
=
\left|
U_{\beta1}^{*}
\,
U_{\alpha1}
+
U_{\beta2}^{*}
\,
U_{\alpha2}
\,
\exp\!\left(
- i
\frac{ \Delta{m}^{2}_{21} \, L }{ 2 \, p }
\right)
\right|^2
+
\left|
\sum_{k=3,4}
U_{{\beta}k}^{*}
\,
U_{{\alpha}k}
\right|^2
\;.
\label{plba2}
\end{eqnarray}
These LBL formulas are derived by
taking into account the fact that 
-- apart from Kam-Land with the MSW solution of the solar neutrino
deficit (see next subsection) --
in LBL experiments
$ \Delta{m}^{2}_{43} L / 2 p \ll 1 $
and dropping the terms proportional to 
the cosines of phases much larger
than $2\pi$
(we have $ \Delta{m}^{2}_{kj} L / 2 p \gg 2\pi $
for $k=3,4$ and $j=1,2$),
which
do not contribute to the oscillation
probabilities averaged over the
neutrino energy spectrum.
The transition probabilities
in scheme B
ensue from the expressions
(\ref{plba1}) and (\ref{plba2})
with the change
\begin{equation}
1 \, , \, 2
\leftrightarrows
3 \, , \, 4
\;.
\label{0721}
\end{equation}

Since scheme B emerges from scheme A by the substitution (\ref{0721}) and
since we will derive bounds on the LBL
oscillation probabilities
$
P^{(\mathrm{LBL,A})}_{\stackrel{\makebox[0pt][l]
{$\hskip-3pt\scriptscriptstyle(-)$}}{\nu_{\alpha}}
\to\stackrel{\makebox[0pt][l]
{$\hskip-3pt\scriptscriptstyle(-)$}}{\nu_{\beta}}}
$
and
$
P^{(\mathrm{LBL,B})}_{\stackrel{\makebox[0pt][l]
{$\hskip-3pt\scriptscriptstyle(-)$}}{\nu_{\alpha}}
\to\stackrel{\makebox[0pt][l]
{$\hskip-3pt\scriptscriptstyle(-)$}}{\nu_{\beta}}}
$
as functions of $A_{\alpha;\beta}$, $c_\alpha$ and $c_\beta$,
it is
evident that such bounds apply equally to both schemes A and B 
and to neutrinos and antineutrinos by
virtue of the definitions (\ref{Aab}) and
(\ref{defca}). Consequently, when dealing with such bounds 
we will omit the superscripts A, B 
indicating the specific scheme.

To derive limits on the
LBL oscillation probabilities
which are given by 
the results of the
SBL oscillation experiments we apply the
Cauchy--Schwarz inequality. It
implies for scheme A that
\begin{equation}
\left|
\sum_{k=1,2}
U_{{\beta}k}
\,
U_{{\alpha}k}^{*}
\,
\exp\!\left(
- i
\frac{ \Delta{m}^{2}_{k1} \, L }{ 2 \, p }
\right)
\right|^2
\leq
c_{\alpha}
\,
c_{\beta}
\;.
\label{111}
\end{equation}
Using this inequality and the definition
(\ref{defca})
of $c_\alpha$,
we find from the LBL probabilities in
Eqs.(\ref{plba1}) and (\ref{plba2})
that the survival probabilities
$P^{(\mathrm{LBL})}_{\nu_{\alpha}\to\nu_{\alpha}}$
and
$P^{(\mathrm{LBL})}_{\bar\nu_{\alpha}\to\bar\nu_{\alpha}}$
are bounded by
\begin{equation}
\left( 1 - c_{\alpha} \right)^2
\leq
P^{(\mathrm{LBL})}_{\stackrel{\makebox[0pt][l]
{$\hskip-3pt\scriptscriptstyle(-)$}}{\nu_{\alpha}}
\to\stackrel{\makebox[0pt][l]
{$\hskip-3pt\scriptscriptstyle(-)$}}{\nu_{\alpha}}}
\leq
c_{\alpha}^2
+
\left( 1 - c_{\alpha} \right)^2
\;.
\label{paa}
\end{equation}
As explained before
these bounds are
scheme-independent. 
In order to obtain bounds on the 
LBL transition probabilities
$P^{(\mathrm{LBL})}_{\nu_{\alpha}\to\nu_{\beta}}$
and
$P^{(\mathrm{LBL})}_{\bar\nu_{\alpha}\to\bar\nu_{\beta}}$
with $\beta\neq\alpha$,
we take into account the definition
(\ref{Aab}) of $A_{\alpha;\beta}$
and the inequality (\ref{111}).
When inserted into Eqs.(\ref{plba1}) and (\ref{plba2})
they imply
\begin{equation}
\frac{1}{4}
\,
A_{\alpha;\beta}
\leq
P^{(\mathrm{LBL})}_{\stackrel{\makebox[0pt][l]
{$\hskip-3pt\scriptscriptstyle(-)$}}{\nu_{\alpha}}
\to\stackrel{\makebox[0pt][l]
{$\hskip-3pt\scriptscriptstyle(-)$}}{\nu_{\beta}}}
\leq
c_{\alpha}
\,
c_{\beta}
+
\frac{1}{4}
\,
A_{\alpha;\beta}
\;.
\label{pab1}
\end{equation}
The bounds (\ref{paa}) and (\ref{pab1})
are the basis of the following
considerations for the different oscillation channels in LBL
experiments.

The smallness of $c_e$ in scheme A
(see Eq.(\ref{A}))
implies that the electron neutrino has a
small mixing with the neutrinos whose mass-squared difference is
responsible for the oscillations of atmospheric and LBL neutrinos
($\nu_1$, $\nu_2$ in scheme A).
Hence, the probability of transitions of atmospheric and
LBL electron neutrinos into other states is suppressed.
Indeed, taking into account the constraint $ c_e \leq a^{0}_{e} $, 
the lower bound on
$P^{(\mathrm{LBL})}_{\bar\nu_{e}\to\bar\nu_{e}}$
and the upper bounds on
$
P^{(\mathrm{LBL})}_{\stackrel{\makebox[0pt][l]
{$\hskip-3pt\scriptscriptstyle(-)$}}{\nu_{\mu}}
\to\stackrel{\makebox[0pt][l]
{$\hskip-3pt\scriptscriptstyle(-)$}}{\nu_{e}}}
$
which we will derive
are rather strict.

Let us discuss first the bounds on
the LBL survival probability
$P^{(\mathrm{LBL})}_{\bar\nu_{e}\to\bar\nu_{e}}$.
With the constraint (\ref{A}) on
$c_{e}$,
Eq.(\ref{paa})
implies that
in both schemes A and B
\begin{equation}
1
-
P^{(\mathrm{LBL})}_{\stackrel{\makebox[0pt][l]
{$\hskip-3pt\scriptscriptstyle(-)$}}{\nu_{e}}
\to\stackrel{\makebox[0pt][l]
{$\hskip-3pt\scriptscriptstyle(-)$}}{\nu_{e}}}
\leq
a^{0}_{e}
\left( 2 - a^{0}_{e} \right)
\;.
\label{081}
\end{equation}
The curve corresponding
to this limit
obtained from the 90\% CL exclusion plot of the Bugey
\cite{Bugey95}
experiment is shown
in Fig. \ref{fig1}
(solid line). For comparison,
the expected sensitivities
of the LBL reactor neutrino experiments
CHOOZ and Palo Verde
are also shown in
Fig. \ref{fig1}
by the dash-dotted and dash-dot-dotted vertical lines, respectively.
These expected sensitivities with respect to 
$1-P^{(\mathrm{LBL})}_{\bar\nu_e\to\bar\nu_e}$
have been extracted by us
from the figures presented in Refs. \cite{CHOOZ,PaloVerde}
showing the sensitivity of the respective experiments in the
two-generation $\sin^{2}2\vartheta$--$\delta{m}^{2}$
plane (here $\vartheta$ is the mixing angle and $\delta{m}^{2}$ is the
mass-squared difference), using the fact that for high values of
$\delta{m}^{2}$ each experiment is sensitive only to the averaged
survival probability
$ P^{(\mathrm{LBL})}_{\bar\nu_e\to\bar\nu_e} = 1 - \frac{1}{2} \sin^{2}2\vartheta $.
Thus,
the vertical lines in Fig. \ref{fig1} correspond to 
$ \frac{1}{2} \sin^2 2\vartheta $ at high $\delta m^2$
in the figures presented in Refs. \cite{CHOOZ,PaloVerde}.
The case of the Kam-Land experiment will be
discussed in sections
\ref{sim}
and
\ref{Matter Corrections}.

Figure \ref{fig1}
shows that,
in the framework of the two schemes (\ref{AB}) with four neutrinos,
which allow to accommodate all 
the indications in favour of neutrino oscillations,
the existing data 
put rather strong limits on the probability of
LBL transitions of
$\nu_e$ into other states
(for
$ \Delta{m}^2 \gtrsim 3 \, \mathrm{eV}^2 $
the upper bound for
$ 1 - P^{(\mathrm{LBL})}_{\bar\nu_{e}\to\bar\nu_{e}} $
is close to the border of the region of sensitivity of the
CHOOZ experiment,
whereas for
$ \Delta{m}^2 \lesssim 3 \, \mathrm{eV}^2 $
it is much smaller).

The shadowed region in Fig. \ref{fig1}
corresponds to the range (\ref{range}) of $\Delta{m}^2$
allowed at 90\% CL by the results of the LSND
and all the other SBL experiments.
It can be seen that
the LSND signal indicates an upper bound for
$1-P^{(\mathrm{LBL})}_{\bar \nu_e \to \bar \nu_e}$
of about
$ 5 \times 10^{-2} $,
smaller than the expected sensitivities of
the CHOOZ and Palo Verde experiments.

Let us stress that,
in the framework of the schemes
under consideration,
the smallness of $c_e$
is a consequence of the solar neutrino problem.
Consider for example scheme A.
The probability of solar neutrinos
to survive is given by
\begin{equation}
P^{(\mathrm{sun},\mathrm{A})}_{\nu_e\to\nu_e}
=
\sum_{k=1,2} |U_{ek}|^4
+
\left( 1 - \sum_{k=1,2} |U_{ek}|^2 \right)^2
P^{(3;4)}_{\nu_e\to\nu_e}
\;,
\label{Psol}
\end{equation}
where $P^{(3;4)}_{\nu_e\to\nu_e}$ is the
survival probability
due to the mixing of $\nu_e$ with $\nu_3$ and $\nu_4$,
depending on the
small mass-squared difference
$\Delta{m}^{2}_{43}$.
From the results of SBL reactor experiments 
it follows that the quantity
$
c_e
\equiv
\displaystyle
\sum_{k=1,2} |U_{ek}|^2
$
can be small or large (close to one).
In order to
have the energy dependence of
the survival probability
$P^{(\mathrm{sun},\mathrm{A})}_{\nu_e\to\nu_e}$
and the suppression of the flux of solar $\nu_e$'s
that are required for the
explanation of the data of solar neutrino experiments,
we must choose a small value of 
$c_e$.
In this case,
the
survival probability of $\bar\nu_e$'s
in LBL reactor experiments
is close to one.

We want to emphasize that from the constraint on $a^0_\mu$ in
Eq.(\ref{A}) and
from Eq.(\ref{paa}) no non-trivial bound on the 
$
\stackrel{\makebox[0pt][l]
{$\hskip-3pt\scriptscriptstyle(-)$}}{\nu_{\mu}}
\to\stackrel{\makebox[0pt][l]
{$\hskip-3pt\scriptscriptstyle(-)$}}{\nu_{\mu}}
$
survival probability can be derived.

Let us now discuss the bounds on
$
\stackrel{\makebox[0pt][l]
{$\hskip-3pt\scriptscriptstyle(-)$}}{\nu_{\mu}}
\to\stackrel{\makebox[0pt][l]
{$\hskip-3pt\scriptscriptstyle(-)$}}{\nu_{e}}
$
transitions in LBL accelerator experiments.
We will compare these bounds with
the expected sensitivities of the
K2K \cite{K2K},
MINOS \cite{MINOS}
and
ICARUS \cite{ICARUS}
experiments.
Taking into account the constraints (\ref{A}) on
$c_{e}$ and (\ref{A0}) on
$A_{\mu;e}$,
Eq.(\ref{pab1})
implies that
in both schemes A and B
\begin{equation}
P^{(\mathrm{LBL})}_{\stackrel{\makebox[0pt][l]
{$\hskip-3pt\scriptscriptstyle(-)$}}{\nu_{\mu}}
\to\stackrel{\makebox[0pt][l]
{$\hskip-3pt\scriptscriptstyle(-)$}}{\nu_{e}}}
\leq
a^{0}_{e}
+
\frac{1}{4}
\,
A_{\mu;e}^{0}
\;.
\label{042}
\end{equation}
Conservation of probability
and Eq.(\ref{paa}) lead to a further upper bound:
\begin{equation}
P^{(\mathrm{LBL})}_{\nu_\alpha \rightarrow \nu_\beta}
\leq
1
-
P^{(\mathrm{LBL})}_{\nu_\alpha \rightarrow \nu_\alpha}
\leq
c_{\alpha} \left( 2 - c_{\alpha} \right)
\qquad (\alpha \neq \beta)\,.
\label{pab2}
\end{equation}
In general
$P^{(\mathrm{LBL})}_{\nu_\beta \rightarrow \nu_\alpha}$
can be different from
$P^{(\mathrm{LBL})}_{\nu_\alpha \rightarrow \nu_\beta}$
(if CP is violated in the lepton sector),
but conservation of probability
gives the same upper bound as
Eq.(\ref{pab2}) for the opposite transition
$\nu_\beta \rightarrow \nu_\alpha$:
\begin{equation}
P^{(\mathrm{LBL})}_{\nu_\beta \rightarrow \nu_\alpha}
\leq
1
-
P^{(\mathrm{LBL})}_{\nu_\alpha \rightarrow \nu_\alpha}
\leq
c_{\alpha} \left( 2 - c_{\alpha} \right)
\qquad (\alpha \neq \beta)\,.
\label{pab3}
\end{equation}
Finally, these two equations hold evidently also for antineutrinos.
Thus from Eq.(\ref{pab3}) and the constraint (\ref{A}) on $c_e$
we obtain
\begin{equation}
P^{(\mathrm{LBL})}_{\stackrel{\makebox[0pt][l]
{$\hskip-3pt\scriptscriptstyle(-)$}}{\nu_{\mu}}
\to\stackrel{\makebox[0pt][l]
{$\hskip-3pt\scriptscriptstyle(-)$}}{\nu_{e}}}
\leq
a^{0}_{e}
\left( 2 - a^{0}_{e} \right) \, .
\label{0421}
\end{equation}
Numerically, this bound is better than the bound (\ref{042}) for the
SBL parameter $\Delta m^2 \lesssim 0.4 \, \mathrm{eV}^2$.

Combining Eqs.(\ref{042}) and (\ref{0421}), we finally arrive at
\begin{equation}
P^{(\mathrm{LBL})}_{\stackrel{\makebox[0pt][l]
{$\hskip-3pt\scriptscriptstyle(-)$}}{\nu_{\mu}}
\to\stackrel{\makebox[0pt][l]
{$\hskip-3pt\scriptscriptstyle(-)$}}{\nu_{e}}}
\leq
\min\!\left(
a^{0}_{e}
\left( 2 - a^{0}_{e} \right)
\, , \,
a^{0}_{e}
+
\frac{1}{4}
\,
A^{0}_{\mu;e}
\right)
\; .
\label{pab12}
\end{equation}
The curve corresponding
to this limit
obtained from the 90\% CL exclusion plots of the Bugey
\cite{Bugey95}
experiment for
$a^{0}_{e}$
and
of the
BNL E734
\cite{BNLE734},
BNL E776
\cite{BNLE776}
and
CCFR
\cite{CCFR97}
experiments
for
$A_{\mu;e}^{0}$
is shown
in Figs. \ref{fig2} and \ref{fig3} by the short-dashed line.
For comparison,
the expected sensitivities
of the LBL accelerator neutrino experiments
K2K \cite{K2K},
MINOS \cite{MINOS}
and
ICARUS \cite{ICARUS} are also indicated
(the dash-dotted vertical line in Fig. \ref{fig2}
and the dash-dotted and dash-dot-dotted
vertical lines in Fig. \ref{fig3}, respectively).
These sensitivities have been obtained
from the figures presented in Refs. \cite{K2K,MINOS,ICARUS}
showing the sensitivities of the respective experiments in the
two-generation $\sin^{2}2\vartheta$--$\delta{m}^{2}$ plane
with the method explained in the context of LBL
reactor experiments.
Note, however, that the short-dashed lines in
Figs. \ref{fig2} and \ref{fig3} have to be corrected for matter
effects. This will be done in the next section.\footnote{The
short-dashed lines in both figures are identical, however, they
receive different matter corrections for K2K on the one hand and
MINOS and ICARUS on the other hand. This will become clear in the
next section.}

The shadowed areas
in Figs. \ref{fig2} and \ref{fig3}
represent the range (\ref{range}).
The LSND \cite{LSND} experiment
also supplies the lower bound in vacuum 
\begin{equation} \label{pmin}
\frac{1}{4}
\,
A_{\mu;e}^{\mathrm{min}}
\leq
P^{(\mathrm{LBL})}_{\stackrel{\makebox[0pt][l]
{$\hskip-3pt\scriptscriptstyle(-)$}}{\nu_{\mu}}
\to\stackrel{\makebox[0pt][l]
{$\hskip-3pt\scriptscriptstyle(-)$}}{\nu_{e}}}
\end{equation}
where
$A_{\mu;e}^{\mathrm{min}}$
is the minimal value of
$A_{\mu;e}$
allowed at
90\% CL by the LSND experiment.
Evidently, $A_{\mu;e}^{\mathrm{min}}$ only exists for the 
range (\ref{range}) and the function 
$\frac{1}{4} A_{\mu;e}^{\mathrm{min}}$ of $\Delta m^2$ 
is represented in Figs. \ref{fig2} and \ref{fig3} by
the left edge of the darkly shadowed regions.
These regions extend to
the right until they reach the bound (\ref{pab12}).

Figs. \ref{fig2} and \ref{fig3}
show that,
in the framework of the schemes under consideration,
the sensitivities of the
MINOS and ICARUS
experiments are considerably better
than the
upper bound (\ref{pab12}) for
$
P^{(\mathrm{LBL})}_{\stackrel{\makebox[0pt][l]
{$\hskip-3pt\scriptscriptstyle(-)$}}{\nu_{\mu}}
\to\stackrel{\makebox[0pt][l]
{$\hskip-3pt\scriptscriptstyle(-)$}}{\nu_{e}}}
$. 
The sensitivity of the
K2K experiment
does not seem to be sufficient to reveal
LBL
$
\stackrel{\makebox[0pt][l]
{$\hskip-3pt\scriptscriptstyle(-)$}}{\nu_{\mu}}
\to\stackrel{\makebox[0pt][l]
{$\hskip-3pt\scriptscriptstyle(-)$}}{\nu_{e}}
$ 
oscillations, but matter
corrections soften the bound (\ref{pab12}), as we will discuss
quantitatively in the next section.
It is interesting to observe
the existence of the lower bound (\ref{pmin})
on this transition probability
that follows from the
LSND results.
However,
this lower bound is valid only in the case of
LBL neutrino oscillations in vacuum.
The corrections due to the matter effects
in LBL experiments
make it disappear (see section \ref{Matter Corrections}).

In vacuum, the right-hand side of Eq.(\ref{0421}) 
is at the same time an upper bound on 
$
P^{(\mathrm{LBL})}_{\stackrel{\makebox[0pt][l]
{$\hskip-3pt\scriptscriptstyle(-)$}}{\nu_{e}}
\to\stackrel{\makebox[0pt][l]
{$\hskip-3pt\scriptscriptstyle(-)$}}{\nu_{\tau}}}
$.  
This is evident from Eq.(\ref{pab2}). 
The bound (\ref{0421}) is quite prone to matter effects.
On the other hand,
the probability of
$
\stackrel{\makebox[0pt][l]
{$\hskip-3pt\scriptscriptstyle(-)$}}{\nu_{\mu}}
\to\stackrel{\makebox[0pt][l]
{$\hskip-3pt\scriptscriptstyle(-)$}}{\nu_{\tau}}
$
transitions
is not constrained by the results
of SBL experiments.

Finally, a further upper bound on
$
P^{(\mathrm{LBL})}_{\stackrel{\makebox[0pt][l]
{$\hskip-3pt\scriptscriptstyle(-)$}}{\nu_{\alpha}}
\to\stackrel{\makebox[0pt][l]
{$\hskip-3pt\scriptscriptstyle(-)$}}{\nu_{\beta}}}
$ 
for $\alpha \neq \beta$ is gained from
Eq.(\ref{pab1}).
Since
$A_{\alpha;\beta} \leq 4(1-c_\alpha)(1-c_\beta)$,
we have
\begin{equation}\label{bbb}
P^{(\mathrm{LBL})}_{\stackrel{\makebox[0pt][l]
{$\hskip-3pt\scriptscriptstyle(-)$}}{\nu_{\alpha}}
\to\stackrel{\makebox[0pt][l]
{$\hskip-3pt\scriptscriptstyle(-)$}}{\nu_{\beta}}}
\leq
c_\alpha c_\beta + (1-c_\alpha)(1-c_\beta)
\qquad
(\alpha\neq\beta)
\;.
\end{equation}
Obviously, if $c_\alpha = c_\beta = 0$ or 1 is in the allowed range of
these quantities,
then this upper bound is 1 and thus is trivial.
This
leaves only
$\alpha = \mu$ and $\beta = e$
as a non-trivial case,
with
\begin{equation}\label{bbb1}
P^{(\mathrm{LBL})}_{\stackrel{\makebox[0pt][l]
{$\hskip-3pt\scriptscriptstyle(-)$}}{\nu_{\mu}}
\to\stackrel{\makebox[0pt][l]
{$\hskip-3pt\scriptscriptstyle(-)$}}{\nu_{e}}}
\leq
a^0_e + a^0_\mu - 2 a^0_e a^0_\mu \, .
\end{equation}
The dotted curves in Figs. \ref{fig2} and \ref{fig3}
show this limit
with $a^{0}_{e}$  and $a^{0}_{\mu}$
obtained from
the 90\% CL exclusion plots
of the Bugey \cite{Bugey95}
$\bar\nu_{e}\to\bar\nu_{e}$
experiment
and of
the CDHS \cite{CDHS84} and CCFR \cite{CCFR84}
$\nu_{\mu}\to\nu_{\mu}$
experiments,
respectively.
For
$ a^0_\mu \ll a^0_e \ll 1 $
this bound is about half of that given by
Eq.(\ref{0421}).
However,
since $a^0_\mu$ is only small
in the same range of $\Delta{m}^2$ where
$A_{\mu;e}^{0}$ is small,
numerically the bound
(\ref{bbb1}),
which is stable against matter effects, 
turns out to be worse than
the bound (\ref{042}) in its matter-corrected form
(see section \ref{Matter Corrections}).

\subsection{The case $\Delta m^2_{\mathrm{sun}}L/2p \sim 1$}
\label{sim}

If the MSW
effect is responsible for the solar neutrino deficit
the phase
\begin{equation}
\eta
\equiv
\frac{\Delta m^2_{\mathrm{sun}}L}{2p}
\label{eta}
\end{equation}
is not necessarily negligible in LBL reactor experiments.
Indeed, we have
\begin{equation}
\eta
\simeq
2.5 \times 10^{-2}
\left( \frac{ L }{ 1 \mathrm{km} } \right)
\label{eta1}
\end{equation}
for
$ \Delta{m}^2_{\mathrm{sun}} \simeq 10^{-5} \, \mathrm{eV}^2 $
and $ p \simeq 1 \mathrm{MeV} $.
Hence, the phase $\eta$ is negligible in the CHOOZ and Palo Verde experiments,
which have a baseline of about 1 km,
but is not negligible in the Kam-Land experiment,
which has a baseline of about 150 km.
From Eqs.(\ref{eta}) and (\ref{eta1})
one can see that the phase $\eta$ is a
function of neutrino energy and is of order 1 for Kam-Land.
For a vacuum oscillation solution of the solar neutrino deficit the
corresponding phase is always negligible in LBL experiments.

In order to derive a bound
on the survival probability
$ P^{(\mathrm{LBL})}_{\bar\nu_e\to\bar\nu_e} $,
it is convenient to 
write it (in scheme A)
as
\begin{eqnarray}\label{PPee}
P^{(\mathrm{LBL,A})}_{\bar\nu_e\to\bar\nu_e}
& = &
\left| |U_{e1}|^2 + |U_{e2}|^2
\exp{\left(-i \frac{\Delta{m}^2_{21}L}{2p}\right)} \right|^2
\nonumber
\\
&& +
\left( |U_{e3}|^2 + |U_{e4}|^2 \right)^2 
- 2 |U_{e3}|^2 \, |U_{e4}|^2
\left( 1-\cos\eta \right)
\,.
\end{eqnarray}
It is clear that this probability is bounded from below by
\begin{eqnarray}
P^{(\mathrm{LBL,A})}_{\bar\nu_e\to\bar\nu_e}
& \geq &
(1-c_e)^2 - 2 |U_{e3}|^2 |U_{e4}|^2 (1-\cos\eta)
\nonumber
\\
& \geq &
(1-c_e)^2 \cos^2 \frac{\eta}{2}
\,,
\end{eqnarray}
where we have used the definition of $c_e$
given in  Eq.(\ref{defca}).
Taking into account the constraint (\ref{A})
on $c_e$, we obtain
\begin{equation}\label{lmsw}
1 - P^{(\mathrm{LBL})}_{\bar\nu_e\to\bar\nu_e}
\leq
1 - \cos^2 \frac{\eta}{2} \left(1-a^0_e\right)^2
\,.
\end{equation}

In the case of a small mixing MSW solution
of the solar neutrino problem,
either $|U_{e3}|^2$ or $|U_{e4}|^2$
is very small and the contribution of the term
$
2 \, |U_{e3}|^2 \, |U_{e4}|^2
\left( 1-\cos\eta \right)
$
in Eq.(\ref{PPee})
is negligible.
Hence, in this case the vacuum bound (\ref{081}) on
$ 1 - P^{(\mathrm{LBL})}_{\bar\nu_e\to\bar\nu_e} $
is valid for all reactor LBL experiments,
including Kam-Land.
This bound is represented by the short-dashed line in Fig. \ref{fig4}
for the Kam-Land experiment.

In the case of a large mixing MSW solution,
the contribution of the term
$
2 \, |U_{e3}|^2 \, |U_{e4}|^2
\left( 1-\cos\eta \right)
$
in Eq.(\ref{PPee})
is not negligible.
It is evident from Eq.(\ref{eta})
that the bound (\ref{lmsw})
depends on the neutrino energy.
For example,
assuming
$ \Delta{m}^2_{\mathrm{sun}} \simeq 10^{-5} \, \mathrm{eV}^2 $,
for $ p = 2, \, 4, \, 7 \, \mathrm{MeV} $
we have,
respectively,
$ \eta \simeq 1.9, \, 0.95, \, 0.54 $
and
$ \cos^2 \frac{\eta}{2} \simeq
0.34 , \,
0.79 , \,
0.93
$
in Kam-Land.
The bounds
derived with Eq.(\ref{lmsw})
corresponding to these neutrino momenta
are represented by the dotted, dash-dotted and dash-dot-dotted lines
in Fig. \ref{fig4}
for the Kam-Land experiment.
One can see that for neutrino energies around 2 MeV
the upper bound for
$ 1 - P^{(\mathrm{LBL})}_{\bar\nu_e\to\bar\nu_e} $
in Kam-Land practically disappears.
Hence a measurement of a large transition probability
$ 1 - P^{(\mathrm{LBL})}_{\bar\nu_e\to\bar\nu_e} $
in the Kam-Land experiment
at neutrino energies around 2 MeV
and a suppression of the same probability
at neutrino energies bigger than about 4 MeV
would be an indirect indication in favour of
the large mixing angle MSW solution
of the solar neutrino problem.

If a large mixing MSW solution of the solar neutrino deficit
is realized in nature,
the value of $\Delta m^2_{\mathrm{sun}}$ could be determined
by an experiment like Kam-Land,
having a sufficient neutrino energy resolution.
Considering
Eq.(\ref{PPee}) and neglecting the first term on the right-hand side
which is suppressed by $(a^0_e)^2$ we obtain
\begin{equation}
P^{(\mathrm{LBL,A})}_{\bar\nu_e\to\bar\nu_e} \simeq
(1-c_e)^2
- 4 \, |U_{e3}|^2 \, |U_{e4}|^2
\sin^2 \frac{\eta}{2}
\,.
\end{equation}
As a function of $p$ this survival probability has maxima at 
\begin{equation}
p_0 = \frac{\Delta m^2_{\mathrm{sun}}L}{2 \pi (2k+1)}
\qquad (k=0,1,2,\dots)
\,.
\end{equation}
A measurement of these maxima would allow the determination of 
$\Delta m^2_{\mathrm{sun}}$.

\section{Matter Corrections}
\label{Matter Corrections}

In this section we will derive
the expressions for the LBL transitions in matter
in the schemes (\ref{AB}) with mixing of four neutrinos.
These schemes contain
active and sterile neutrinos.
In such a case, in the effective
Hamiltonian of the interaction of neutrinos with matter there is an
additional neutral current term apart from the usual charged current
term.
For the total Hamiltonians of neutrinos and antineutrinos we have
the following expressions
in the flavour representation \cite{MSW}:
\begin{equation}\label{Hnu}
H_{\nu} = \frac{1}{2p} \left( U \hat{M}^2 U^\dagger + 
\mathrm{diag}\, (a_{CC}, 0, 0, a_{NC}) \right) \, ,
\end{equation}
\begin{equation}\label{Hanu}
H_{\bar\nu} = \frac{1}{2p} \left( U^* \hat{M}^2 U^T - 
\mathrm{diag}\, (a_{CC}, 0, 0, a_{NC}) \right) \, .
\end{equation}
Here we have defined
\begin{eqnarray}
&&
a_{CC}
=
2\sqrt{2} \, G_F \, N_e \, p
\simeq
2.3 \times 10^{-4} \, \mathrm{eV}^2
\left( \frac{ \rho }{ 3 \, \mathrm{g} \, \mathrm{cm}^{-3} } \right)
\left( \frac{ p }{ 1 \, \mathrm{GeV} } \right)
\;,
\label{acc}
\\
&&
a_{NC}
=
\sqrt{2} \, G_F \, N_n \, p
\simeq
\frac{1}{2} \, a_{CC} \;,
\label{anc}
\end{eqnarray}
$\hat{M}^2$ denotes the diagonal matrix of the squares of the neutrino masses,
$G_F$ is the Fermi constant,
$N_e$ and $N_n$ are the electron and neutron number
density,
respectively,
and
$\rho$ is the density of matter,
which in the Earth's crust is on average
$ 3 \, \mathrm{g} \, \mathrm{cm}^{-3} $.
Since the lithosphere consists mainly of elements where the neutron
number equals the proton number,
we have
$N_e \simeq N_n \simeq
\frac{N_{\mathrm{A}}}{2}
\frac{\rho}{1\mathrm{g}}
$,
where
$N_{\mathrm{A}}$
is the Avogadro number.
The parameters $a_{CC}$ and $a_{NC}$
can be as large as
$\Delta m^2_{\mathrm{atm}}$,
which is relevant for LBL oscillations.
Their
effects on the bounds for transition probabilities in LBL experiments
need not be
negligible, as we shall see.

In order to assess the size of
matter effects,
we consider the simplifying 
approximation of constant $N_e$ and $N_n$,
which is rather good
in the case of LBL experiments.
Furthermore,
in the following we
will concentrate on the scheme A
and we will consider only the neutrino 
Hamiltonian (\ref{Hnu}).
At the end of this section we will see
that,
as in the vacuum case,
the bounds obtained for neutrinos in the scheme A
are also valid
for antineutrinos in scheme A
and for neutrinos and antineutrinos in scheme B.

In order to obtain
the expressions for the transition probabilities of
neutrinos in matter,
we have to diagonalize the Hamiltonian (\ref{Hnu}).
With the diagonalization matrix $U'$ we have
\begin{equation}
H_\nu = U' \frac{\hat{\epsilon}}{2p} U'^\dagger
\end{equation}
where
$ \hat{\epsilon} = \mathrm{diag}(\epsilon_1,\ldots,\epsilon_4) $
and
$ U'^\dagger U' = \mathbf{1} $.
Note that in the vacuum case
$U'=U$ and
$ \epsilon_j = m_j^2 $
$(j=1,\ldots,4)$.
It is convenient to use the basis where $H_0 \equiv H_\nu
(a_{CC}=a_{NC}=0)$ is diagonal.
In this basis we have
\begin{equation}
\hat{H}_\nu \equiv U^\dagger H_\nu U =
\frac{1}{2p} \left( \hat{M}^2 + 
U^\dagger \mathrm{diag}\, (a_{CC}, 0, 0, a_{NC}) U \right) =
\frac{1}{2p} R \hat{\epsilon} R^\dagger
\end{equation}
and the unitary matrix $U'$ is given by
\begin{equation}
U' = UR \, .
\end{equation}
Since $a_{CC} \ll \Delta m^2$,
where
$ \Delta m^2 \equiv m_4^2-m_1^2 \sim 1 $ eV$^2$
is relevant for SBL oscillations,
it is obvious that,
apart from corrections of order $a_{CC}/\Delta m^2$,
the matrix
$R$ has the block structure
\begin{equation}\label{block}
R = \left( \begin{array}{cc}    
R_{\mathrm{atm}} & 0 \\ 0 & R_{\mathrm{sun}} \end{array} \right),
\end{equation}
where $R_{\mathrm{atm}}$ and $R_{\mathrm{sun}}$ are 
2$\times$2 unitary matrices. All our considerations in this paper are
based on this approximation. A
glance at Tab. \ref{tab1} shows that the ratio $a_{CC}/\Delta m^2$
is less than $10^{-2}$
for all the LBL experiments of the first generation.
The block structure of $R$ has the consequence that
\begin{equation}\label{scalarp}
\sum_{j=1,2} U'_{\alpha j}U'^*_{\beta j} =
\sum_{j=1,2}  U_{\alpha j} U^*_{\beta j}
\end{equation}
and therefore
\begin{equation}\label{ca}
c_\alpha = \sum_{j=1,2} |U_{\alpha j}|^2 = 
\sum_{j=1,2} |U'_{\alpha j}|^2 \, .
\end{equation}

It is easy to calculate the energy
eigenvalues
up to terms of order $a_{CC}/\Delta m^2$.
We are interested in
the differences $\epsilon_2 - \epsilon_1$ and
$\epsilon_4 - \epsilon_3$.
Defining
\begin{equation}
|U_{\alpha 1}|^2 = c_\alpha \cos^2 \gamma_\alpha \quad \mbox{and} \quad
|U_{\alpha 2}|^2 = c_\alpha \sin^2 \gamma_\alpha \quad \mbox{for} \quad
\alpha = e, s
\end{equation}
and
\begin{equation}
\delta = \arg \, (U_{e1} U^*_{e2} U^*_{s1} U_{s2}) \, ,\quad
y_e = \frac{a_{CC} c_e}{\Delta m^2_{21}}\, ,\quad
y_s = \frac{a_{NC} c_s}{\Delta m^2_{21}}
\end{equation}
we obtain for the atmospheric neutrino sector
\begin{equation}
\epsilon_2 - \epsilon_1 = 
\Delta m^2_{21} \left[ \left| 1-y_e
\mathrm{e}^{2i\gamma_e} - y_s \mathrm{e}^{2i\gamma_s} \right|^2 -
4 y_e y_s \sin 2\gamma_e \sin 2\gamma_s \sin^2 \frac{\delta}{2}
\right]^{1/2} \, .
\end{equation}
Similarly, with
\begin{equation}
|U_{\alpha 3}|^2 = (1-c_\alpha) \cos^2 \gamma'_\alpha \quad \mbox{and} \quad
|U_{\alpha 4}|^2 = (1-c_\alpha) \sin^2 \gamma'_\alpha \quad \mbox{for} \quad
\alpha = e, s
\end{equation}
and
\begin{equation}
\delta' = \arg \, (U_{e3} U^*_{e4} U^*_{s3} U_{s4}) \, ,\quad 
y'_e = \frac{a_{CC} (1-c_e)}{\Delta m^2_{43}}\, ,\quad
y'_s = \frac{a_{NC} (1-c_s)}{\Delta m^2_{43}}
\end{equation}
we have for the solar sector
\begin{equation}\label{e43}
\epsilon_4 - \epsilon_3 =
\Delta m^2_{43} \left[ \left| 1-y'_e
\mathrm{e}^{2i\gamma'_e} - y'_s \mathrm{e}^{2i\gamma'_s} \right|^2 -
4 y'_e y'_s \sin 2\gamma'_e \sin 2\gamma'_s \sin^2 \frac{\delta'}{2}
\right]^{1/2} \, .
\end{equation}

Looking at Eq.(\ref{acc}) or Tab. \ref{tab1} it can be read off that $a_{CC}$ is
not necessarily
smaller than $\Delta m^2_{21}$ relevant for
LBL neutrino oscillations and thus $\epsilon_2 - \epsilon_1$
could be different from
$\Delta m^2_{21}$. 
Since $\Delta m^2_{43}$ is of the
order $10^{-5}$ or $10^{-10}$ eV$^2$, in the solar sector 
$a_{CC}$ is much larger
than the relevant mass-squared difference $\Delta m^2_{\mathrm{sun}}$
except for reactor experiments with the MSW 
solution of the solar neutrino puzzle (see Tab. \ref{tab1}). 
In any case,
$\epsilon_4 - \epsilon _1 \simeq \Delta m^2$
holds and therefore
the LBL oscillation probabilities
averaged over the fast oscillations due to $\Delta{m}^2$
are given by
\begin{equation}\label{P}
P^{(\mathrm{LBL,A})}_{\nu_\alpha\to\nu_\beta} =
\left| \sum_{j=1,2} U'_{\beta j}U'^*_{\alpha j}
\exp{\left(-i \frac{\epsilon_j}{2p} \, L\right)} \right|^2 +
\left| \sum_{k=3,4} U'_{\beta k}U'^*_{\alpha k}
\exp{\left(-i \frac{\epsilon_k}{2p} \, L\right)} \right|^2\, ,
\end{equation}
analogously to the vacuum case (see Eqs.(\ref{plba1}) and (\ref{plba2})).

A first upper bound on $P^{\mathrm{(LBL)}}_{\nu_\alpha \to \nu_\beta}$
is obtained by simply applying the Cauchy--Schwarz inequality to
Eq.(\ref{P}). Taking into account Eq.(\ref{ca}) this leads
to
\begin{equation}\label{CSb}
P^{(\mathrm{LBL})}_{\nu_\alpha \to \nu_\beta} \leq 
c_\alpha c_\beta + (1-c_\alpha)(1-c_\beta).
\end{equation}
It is remarkable that with Eq.(\ref{CSb})
we have recovered Eq.(\ref{bbb}) of
the vacuum case.
The present discussion reveals that this
equation is correct in matter apart
from terms of order $a_{CC}/\Delta m^2$.
Eq.(\ref{CSb}) is also valid for
antineutrinos because the
unitary matrix which diagonalizes $H_{\bar\nu}$ 
also has the block structure (\ref{block}) and in Eq.(\ref{ca}) the
phases of $U$ do not enter.

To derive matter corrections to the other bounds developed in
section \ref{Vacuum Bounds} it is convenient to 
write Eq.(\ref{P}) in the form
\begin{equation}\label{PP}
P^{(\mathrm{LBL,A})}_{\nu_\alpha \to \nu_\beta} =
P'_{\nu_\alpha \to \nu_\beta}
- 2 \, \mathrm{Re}
\left[ U'_{\alpha 3}U'^*_{\beta 3}U'^*_{\alpha 4}U'_{\beta 4}
\left( 1-\exp{(-i\omega)} \right) \right]
\end{equation}
where
\begin{equation}\label{P'}
P'_{\nu_\alpha\to\nu_\beta} = 
\left| U'_{\beta 1}U'^*_{\alpha 1} + U'_{\beta 2}U'^*_{\alpha 2}
\exp{\left(-i \frac{\epsilon_2 - \epsilon_1}{2p} \, L\right)} \right|^2 +
\left| U'_{\beta 3}U'^*_{\alpha 3} + U'_{\beta 4}U'^*_{\alpha 4}
\right|^2 
\end{equation}
and
\begin{equation}\label{omega}
\omega \equiv \frac{\epsilon_4 - \epsilon_3}{2p}L \, .
\end{equation}

The Cauchy--Schwarz inequality and Eqs.(\ref{scalarp}) and (\ref{ca}) allow 
to bound the expression Eq.(\ref{P'}) for $\alpha \neq \beta$ by
\begin{equation}\label{CS}
\frac{1}{4} A_{\alpha;\beta} \leq 
P'_{\nu_\alpha \to \nu_\beta}
\leq c_\alpha c_\beta + \frac{1}{4} A_{\alpha;\beta} \, .
\end{equation}
For $\alpha = \beta$ we have instead
\begin{equation}\label{CS1}
(1-c_\alpha)^2 \leq
P'_{\nu_\alpha \to \nu_\alpha} 
\leq c_\alpha^2 + (1-c_\alpha)^2 \, .
\end{equation}
These two equations are the analogues of Eqs.(\ref{pab1}) and
(\ref{paa}), respectively.
Matter corrections are characterized by the parameter $\omega$.
For $\omega = 0$ the vacuum bounds on 
$P^{(\mathrm{LBL})}_{\nu_\alpha \to \nu_\beta}$ ensue. 
Inspection of Eq.(\ref{e43}) shows that,
taking into account Eqs.(\ref{acc}) and (\ref{anc}),
if $y'_{e,s} \gg 1$
the maximal value of the parameter
$\omega$ is given by
\begin{equation}
\omega_{\mathrm{max}} \simeq \frac{3}{2} \, \frac{a_{CC}L}{2p}
=
\frac{3}{2}
\sqrt{2} \, G_F \, N_e \, L
=
8.6 \times 10^{-4}
\left( \frac{ L }{ 1 \mathrm{km} } \right)
\label{omax}
\end{equation}
for
$ \rho = 3 \, \mathrm{g} \, \mathrm{cm}^{-3} $.
Note that in this case
$\omega_{\mathrm{max}}$
does not depend on the neutrino energy,
but only on the propagation distance $L$.
Hence,
the matter corrections to the bounds
for the LBL transition probabilities
that we will derive
in the case $y'_{e,s} \gg 1$
are independent from the neutrino energy
and the corresponding bounds
apply to all the energy spectrum of LBL experiments.
From Eq.(\ref{omax})
one can see that these matter corrections
could be relevant for
$ L \gtrsim 100 \, \mathrm{km} $.
The size of $\omega_{\mathrm{max}}$ in the individual LBL experiments
can be looked up in Tab. \ref{tab1}.
This parameter is not small for the
MINOS and ICARUS experiments.
In Tab. \ref{tab1}
the CHOOZ and Palo Verde experiments are not listed because
their respective neutrino beams do not propagate in matter.
Anyway, for baselines around 1 km matter effects are totally
negligible,
as can be seen from
Eqs.(\ref{PP})--(\ref{omax}).

The condition $y'_{e,s} \gg 1$ is satisfied
in all the LBL experiments of the first generation if
$\Delta{m}^2_{\mathrm{sun}}$ is either in the range of the MSW
or of the vacuum oscillation solution of the solar neutrino problem,
apart from the LBL reactor experiments if the MSW
effect is responsible for the solar neutrino deficit. 
In this case we have $\Delta{m}^2_{\mathrm{sun}} \gg a_{CC}$,
which implies that
$y'_{e,s} \ll 1$
and
$ \omega \simeq \eta $
given in Eq.(\ref{eta}).
Furthermore,
since in this case $a_{CC}$
is much smaller than all the $\Delta{m}^2$'s,
the mixing matrix in matter
is the same as in vacuum and the oscillations in LBL reactor experiments
are the same as in vacuum.
Hence, this case coincides with that discussed
in section \ref{sim}.

For the same reasons as in the case of
Eq.(\ref{CSb}), the two inequalities (\ref{CS}) and (\ref{CS1}) also hold for
antineutrinos. 
The oscillation probabilities for scheme B follow from Eq.(\ref{P}) with
the substitution of indices
\begin{equation}
1 \, , \, 2
\leftrightarrows
3 \, , \, 4
\;.
\label{indices}
\end{equation}
Since the conditions (\ref{B}) for scheme B emerge from
the corresponding conditions (\ref{A}) through the same substitution
(\ref{indices}),
all the bounds derived for the scheme A hold likewise for the scheme B.

\subsection{A bound on $P^{(\mathrm{LBL})}_{\nu_\mu \to \lowercase{\nu_e}}$ 
stable against matter effects}

Repeating the discussion of section \ref{Vacuum Bounds} for completeness, 
the bound (\ref{CSb}) is non-trivial for
the case of $\nu_\mu \to \nu_e$ transitions and thus we obtain
\begin{equation}\label{CSee}
P^{(\mathrm{LBL})}_{\nu_\mu \to \nu_e} \leq a^0_e + a^0_\mu 
- 2\, a^0_e a^0_\mu \, .
\end{equation}
As argued before it holds for both schemes A and B and for 
neutrinos and antineutrinos.
In Figs. \ref{fig2} and \ref{fig3} the upper bound (\ref{CSee}) 
is represented by the dotted curve.

\subsection{The bound on 
$P^{(\mathrm{LBL})}_{\lowercase{\nu_e} \to \lowercase{\nu_e}}$}

From Eqs.(\ref{PP}) and (\ref{CS1})
with $\alpha = \beta = e$
we get the lower bound
\begin{eqnarray}
P^{(\mathrm{LBL,A})}_{\nu_e\to\nu_e}
& \geq &
(1-c_e)^2 - 2 |U'_{e3}|^2 |U'_{e4}|^2 (1-\cos \omega )
\nonumber
\\
& \geq &
(1-c_e)^2 \cos^2 \frac{\omega}{2}
\,.
\end{eqnarray}
Finally, with Eq.(\ref{A}) we arrive at the result
\begin{equation}\label{eb}
1 - P^{(\mathrm{LBL})}_{\nu_e\to\nu_e}
\leq
1 - \cos^2 \frac{\omega}{2} \left(1-a^0_e\right)^2
\,.
\end{equation}
In the approximation $y'_{e,s} \gg 1$ the parameter $\omega$ is equal for
neutrinos and antineutrinos. Therefore, like the bound (\ref{CSee}),
Eq.(\ref{eb}) holds for both schemes and for neutrinos and
antineutrinos.

Taking into account the value
of $\omega_{\mathrm{max}}$
for the Kam-Land experiment
in the case of a vacuum oscillation solution
of the solar neutrino problem
(see Tab. \ref{tab1}),
this equation shows that
matter effects are not negligible in establishing the
upper bound for
$1-P^{(\mathrm{LBL})}_{\bar\nu_e\to\bar\nu_e}$
in this experiment.
This upper bound
is shown by the solid line in Fig. \ref{fig4}
and one can see that there is a small
deviation from the vacuum bound (short-dashed line)
for small values of $\Delta{m}^2$.
However,
the vacuum bound is not substantially modified and
the bound in matter is well below
the sensitivity of the Kam-Land experiment
(represented by vertical long-dashed line).
Hence,
in the case of a vacuum oscillation solution
of the solar neutrino problem
the sensitivity of the Kam-Land experiment
is not enough to observe
$\bar\nu_e\to\bar\nu_e$
in the framework of the 4-neutrino schemes (\ref{AB}).

Conservation of probability leads to
\begin{equation}\label{emb}
P^{(\mathrm{LBL})}_{\nu_e \to \nu_\mu} \leq
1-P^{(\mathrm{LBL})}_{\nu_e \to \nu_e}
\end{equation}
and therefore Eq.(\ref{eb}) also bounds $P^{(\mathrm{LBL})}_{\nu_e \to
\nu_\mu}$.
In LBL accelerator experiments, the initial neutrinos are $\nu_\mu$'s
and not $\nu_e$'s.
If, however, CP is conserved in the lepton sector,
it follows from the CPT theorem that
the transition probabilities in vacuum are invariant under time reversal,
i.e.
$P_{\nu_\alpha\to\nu_\beta} =
P_{\nu_\beta\to\nu_\alpha}$.
In this case, the probabilities of
$\nu_\alpha \to \nu_\beta$ and $\nu_\beta \to \nu_\alpha$
transitions are equal
also in matter if the matter density is symmetric along the neutrino path.
In general,
however,
these two probabilities are different.
Nevertheless,
from conservation of probability we obtain
\begin{equation}\label{abba}
P^{(\mathrm{LBL})}_{\nu_\mu \to \nu_e}
\leq
1 - P^{(\mathrm{LBL})}_{\nu_e \to \nu_e}
\leq
1 - \cos^2 \frac{\omega}{2} \left(1-a^0_e\right)^2
\,.
\end{equation}
Eq.(\ref{abba}) is the matter-corrected version of
Eq.(\ref{0421}).
For the K2K
experiment $\sin^2 \omega_{\mathrm{max}} \simeq 10^{-2}$ is
small, but for the ICARUS and MINOS experiments we obtain
$\sin^2 \omega_{\mathrm{max}} \simeq 0.09$ and therefore, in the
case of these two experiments, matter effects are considerable 
in the bound (\ref{abba}) on the probabilities of LBL
$\nu_\mu \to \nu_e$ transitions. In Figs. \ref{fig2} and \ref{fig3}
the long-dashed lines show the bound (\ref{abba}) for the K2K
experiment and the ICARUS and MINOS experiments, respectively.

Note that the right-hand side of Eq.(\ref{eb}) also
constitutes an upper bound for
$P^{(\mathrm{LBL})}_{\nu_e \to \nu_\tau}$.

\subsection{The upper bound on 
$P^{(\mathrm{LBL})}_{\nu_\mu \to \nu_{\lowercase{e}}}$}

Now we come to matter corrections to the bound (\ref{042})
on $P^{(\mathrm{LBL})}_{\nu_\mu \to \nu_e}$ which
incorporates information on $A_{\mu;e}$. 
The starting point is given by Eqs.(\ref{PP}) and (\ref{CS}). 
To proceed further 
it is necessary to develop a parameterization for $U'_{\alpha j}$.
Without loss of generality we can write
\begin{equation}
U'_{ej} = \sqrt{1-c_e} \, e^{(1)}_{j-2}
\qquad
\mbox{for}
\qquad j=3,4
\,,
\end{equation}
with the orthonormal basis
\begin{equation}\label{ON}
e^{(1)}(\theta) = (\cos \theta, \sin \theta)
\;,
\qquad
e^{(2)}(\theta) = (-\sin \theta, \cos \theta)
\;.
\end{equation}
We expand $U'_{\mu j}$
with $j=3,4$
with respect to this
basis as
\begin{equation}\label{Ube}
U'_{\mu j}
=
\sqrt{1-c_\mu} \sum_{\rho=1,2} p_\rho \, e^{(\rho)}_{j-2}
\;, 
\end{equation}
where
$p_1$ and $p_2$
are complex coefficients
such that 
\begin{equation}\label{norm}
\sum_{\rho=1,2} |p_\rho|^2 = 1
\;.
\end{equation}
Using this parameterization,
from Eqs.(\ref{Aab}) and (\ref{scalarp})
we obtain
\begin{equation}\label{Ame}
A_{\mu;e} = 4(1-c_e)(1-c_\mu) |p_1|^2 \, .
\end{equation}
With these equations we eliminate $|p_1|$ and $|p_2|$ and defining $\sigma
= \arg \, (p^*_1 p_2)$ we arrive at
\begin{eqnarray}
P^{(\mathrm{LBL})}_{\nu_\mu \to \nu_e} & \leq & 
           c_e c_\mu + \frac{1}{4}A_{\mu;e} + 
       (1-\cos \omega) \Bigg\{ \left[ \frac{1}{2} (1-c_e)(1-c_\mu) -
    \frac{1}{4}A_{\mu;e} \right] \sin^2 2\theta - \nonumber\\
 & - &
\frac{1}{8} \sqrt{A_{\mu;e} \left[ 4(1-c_e)(1-c_\mu) - A_{\mu;e}
    \right]} \,\sin 4\theta \cos \sigma \Bigg\} - \nonumber\\
 & - & \frac{1}{4} \sin \omega \sqrt{A_{\mu;e} \left[ 4(1-c_e)(1-c_\mu) 
    - A_{\mu;e}
    \right]} \,\sin 2\theta \sin \sigma \, . \label{meb1}
\end{eqnarray}
Since we do not have information on
the values of $\theta$ and $\sigma$,
we have to maximize
the right-hand side of Eq.(\ref{meb1}).
The maximum with respect to $\sigma$ is easily found:
the maximum of a
function $a \cos \sigma + b \sin \sigma$ with constant $a$ and $b$ is
given by $\sqrt{a^2+b^2}$.
It remains to find the maximum with respect
to $\sin^2 2\theta$ for the resulting function. One can show that,
if $ \cos\omega \geq 0 $,
the maximum
is given by $\sin^2 2\theta = 1$.
This is the case for the K2K, MINOS and ICARUS experiments
because $ \omega_{\mathrm{max}} < \pi/2$
and therefore the following bound applies:
\begin{eqnarray}\label{omeb}
P^{(\mathrm{LBL})}_{\nu_\mu \to \nu_e} & \leq & c_e c_\mu + 
\frac{1}{4} \cos \omega\, A_{\mu;e} + 
\frac{1}{2} (1-\cos \omega)\, (1-c_e)\,(1-c_\mu) +
\nonumber\\
& + & \frac{1}{4} \sin \omega\, \sqrt{A_{\mu;e} \left[
4(1-c_e)(1-c_\mu)-A_{\mu;e} \right]} \, .
\end{eqnarray}
In this inequality
it is difficult to take into account analytically
the conditions (\ref{A}), (\ref{A0}) and
$ \omega \leq \omega_{\mathrm{max}} $.
Hence,
we have done it numerically 
and the result is shown by the
solid curves
in Figs. \ref{fig2} and \ref{fig3}.
In both figures the solid
curve is the most stringent bound on
$
\stackrel{\makebox[0pt][l]
{$\hskip-3pt\scriptscriptstyle(-)$}}{\nu_{\mu}}
\to\stackrel{\makebox[0pt][l]
{$\hskip-3pt\scriptscriptstyle(-)$}}{\nu_{e}}
$
transitions with matter effects. These two solid curves
belong to the main results of this paper.
It is interesting to compare the solid curves in Figs. \ref{fig2}
and \ref{fig3} with the short-dashed lines which
represent the corresponding bounds in vacuum.
For the K2K experiment the solid line deviates from the short-dashed
line only at $\Delta m^2$ close to $0.3 \, \mathrm{eV}^2$, the
lower edge of the range (\ref{range}). The same is true for the
MINOS and ICARUS experiments except that the deviation starts
at larger $\Delta m^2$ values and is more pronounced at the
lower edge of the shadowed area.

As in the previous sections,
the bound (\ref{omeb})
is valid for
neutrinos and antineutrinos.
Although the parameters $\theta$ and
$\sigma$ are in principle different for neutrinos and antineutrinos the
maximization procedure wipes out any difference in the bounds.

\subsection{The lower bound on 
$P^{(\mathrm{LBL})}_{\nu_\mu \to \lowercase{\nu_e}}$}

Since the LSND experiment has seen a positive signal for 
$P^{(\mathrm{LBL})}_{\stackrel{\makebox[0pt][l]
{$\hskip-4pt\scriptscriptstyle(-)$}}{\nu_\mu}
\to\stackrel{\makebox[0pt][l]
{$\hskip-4pt\scriptscriptstyle(-)$}}{\nu_e}}$,
this experiment provides a lower bound
for the amplitude of
$
\stackrel{\makebox[0pt][l]
{$\hskip-4pt\scriptscriptstyle(-)$}}{\nu_\mu}
\to\stackrel{\makebox[0pt][l]
{$\hskip-4pt\scriptscriptstyle(-)$}}{\nu_e}
$
transitions (see Eq.(\ref{pmin})).
Using analogous steps as in the previous section one derives
\begin{eqnarray}\label{lmeb}
P^{(\mathrm{LBL})}_{\nu_\mu \to \nu_e} & \geq & 
\frac{1}{4} \cos \omega\, A_{\mu;e} + 
\frac{1}{2} (1-\cos \omega)\, (1-c_e)\,(1-c_\mu)
\nonumber\\
&&
- \frac{1}{4} \sin \omega\, \sqrt{A_{\mu;e} \left[
4(1-c_e)(1-c_\mu)-A_{\mu;e} \right]} \, .
\end{eqnarray}
Now we have to minimize the right-hand side of Eq.(\ref{lmeb}) with
respect to $c_e$ and $c_\mu$ with the bounds $(\ref{A})$. This procedure
leads to the following result:
\begin{equation}\label{lmeb1}
P^{(\mathrm{LBL})}_{\nu_\mu \to \nu_e} \geq 0 \quad \mbox{for} \quad 
A_{\mu;e} \leq 2\, (1-\cos \omega) a^0_\mu
\end{equation}
and
\begin{eqnarray}
&&
P^{(\mathrm{LBL})}_{\nu_\mu \to \nu_e} \geq
\frac{1}{4} \cos\omega A_{\mu;e} +
\frac{1}{2} (1-\cos \omega) a^0_\mu - \frac{1}{4} \sin \omega
\sqrt{A_{\mu;e} (4a^0_\mu - A_{\mu;e})} \nonumber\\
&&
\mbox{for} \quad 
A_{\mu;e} \geq 2\, (1-\cos \omega) a^0_\mu \, . \label{lmeb2}
\end{eqnarray}
Eq.(\ref{lmeb1}) states that $A_{\mu;e}$ has to be
sufficiently large otherwise the non-trivial lower bound in vacuum
($\omega = 0$) becomes trivial. 

Taking $\omega = 0.63$, the maximal value of the parameter $\omega$
for the ICARUS and MINOS
experiments (see Tab. \ref{tab1}),
the condition for a trivial lower bound on 
$P^{(\mathrm{LBL})}_{\nu_\mu \to \nu_e}$ is given by $A_{\mu;e}
\lesssim 0.3 \, a^0_\mu$. 
Looking at $A_{\mu;e}^{\mathrm{min}}$ of the LSND
experiment and the function $a^0_\mu$ 
one sees that indeed this condition is fulfilled. Thus 
matter effects make the non-trivial
lower bound of the vacuum case disappear. For the K2K experiment
the analogous condition for triviality is given by
$A_{\mu;e} \lesssim 0.04 \, a^0_\mu$.
Here the triviality condition is not fulfilled but
K2K does not seem to have sufficient sensitivity to reach small
enough $P^{(\mathrm{LBL})}_{\nu_\mu \to \nu_e}$. However, for such small
oscillation probabilities it would be necessary to take corrections of
order $a_{CC}/\Delta m^2$ into account. Thus also in this case
the lower bound seems to be irrelevant and
the shadowed areas (dark and light) in Figs. \ref{fig2}
and \ref{fig3} show the allowed regions for
$
\stackrel{\makebox[0pt][l]
{$\hskip-4pt\scriptscriptstyle(-)$}}{\nu_\mu}
\to\stackrel{\makebox[0pt][l]
{$\hskip-4pt\scriptscriptstyle(-)$}}{\nu_e}
$
transitions taking into account matter effects for the K2K
experiment (Fig. \ref{fig2}) and the MINOS and ICARUS experiments
(Fig. \ref{fig3}). 

\section{Three Massive Neutrinos}
\label{Three Massive Neutrinos}

It is worthwhile to have a look at LBL
neutrino oscillation experiments
neglecting some of the
present hints for neutrino oscillations.
It is possible that
not all these hints
will be substantiated in the course of time and it is
useful to check which features
are actually dependent on or independent 
from them.

In this section we consider the minimal scenario
of mixing of three neutrinos.
We will
assume that of the two differences
of squares of neutrino masses one is
relevant for SBL
oscillations
and the other one for LBL
oscillations
(see also Refs. \cite{TA96,AR96,MN96}).
Hence,
in this section
we adopt the point of view 
that not neutrino mixing but other
reasons could explain the solar neutrino data.
With these assumptions,
there are two possible 3-neutrino mass spectra:
\begin{equation}
(\mbox{\textrm{I}})
\qquad
\underbrace{
\overbrace{m_1 < m_2}^{\mathrm{LBL}}
\ll
m_3
}_{\mathrm{SBL}}
\qquad \mbox{and} \qquad 
(\mbox{\textrm{II}})
\qquad
\underbrace{
m_1
\ll
\overbrace{m_2 < m_3}^{\mathrm{LBL}}
}_{\mathrm{SBL}}
\;.
\end{equation}
In both schemes
\textrm{I} and \textrm{II},
$\Delta m^2_{31}$
is assumed to be relevant for
neutrino oscillations in SBL experiments.
In this case,
the SBL oscillation probabilities
depend on
$|U_{e3}|^2$
and
$|U_{\mu3}|^2$
in the scheme \textrm{I}
\cite{BBGK}
and
on
$|U_{e1}|^2$
and
$|U_{\mu1}|^2$
in the scheme \textrm{II}
\cite{BGKP}.
There are three regions
of these quantities
which are
allowed by the results
of disappearance experiments
(see Refs. \cite{BBGK,BGKP}):
\begin{equation}\label{regions}
\begin{array}{lll} \displaystyle
(1) \qquad &
|U_{ek}|^2 \geq 1-a^0_e
\;,
\qquad
&
|U_{{\mu}k}|^2 \leq a^0_\mu
\;,
\\[3mm] \displaystyle
(2) \qquad &
|U_{ek}|^2 \leq a^0_e
\;,
\qquad
&
|U_{\mu k}|^2 \leq a^0_\mu
\;,
\\[3mm] \displaystyle
(3) \qquad &
|U_{ek}|^2 \leq a^0_e
\;,
\qquad
&
|U_{\mu k}|^2 \geq 1-a^0_\mu
\;,
\end{array} 
\end{equation}
with
$k=3$ for the scheme \textrm{I}
and
$k=1$ for the scheme \textrm{II}\footnote{
For a comparison,
the schemes \textrm{I}, \textrm{II}
and the regions 1, 2, 3 are
called hierarchies II, I and regions A, B, C, respectively, 
in Ref. \cite{TA96}.}
(for the definition of $a^0_e$ and $a^0_\mu$
see Eq.(\ref{a0})).

The neutrino and antineutrino
LBL oscillation probabilities
in scheme \textrm{I}
are given by
\begin{eqnarray}
&&
P^{(\mathrm{LBL},\mathrm{I})}_{\nu_\alpha\to\nu_\beta}
=
\left|
U_{\beta1}
\,
U_{\alpha1}^{*}
+
U_{\beta2}
\,
U_{\alpha2}^{*}
\,
\exp\!\left(
- i
\frac{ \Delta{m}^{2}_{21} \, L }{ 2 \, p }
\right)
\right|^2
+
|U_{{\beta}3}|^2
\,
|U_{{\alpha}3}|^2
\;,
\label{0171}
\\
&&
P^{(\mathrm{LBL},\mathrm{I})}_{\bar\nu_\alpha\to\bar\nu_\beta}
=
\left|
U_{\beta1}^{*}
\,
U_{\alpha1}
+
U_{\beta2}^{*}
\,
U_{\alpha2}
\,
\exp\!\left(
- i
\frac{ \Delta{m}^{2}_{21} \, L }{ 2 \, p }
\right)
\right|^2
+
|U_{{\beta}3}|^2
\,
|U_{{\alpha}3}|^2
\;.
\label{0172}
\end{eqnarray}
The transition probabilities
in the scheme \textrm{II}
can be obtained from the expressions
(\ref{0171}) and (\ref{0172})
with the cyclic permutation of the indices
\begin{equation}
1 \, , \, 2 , \, 3
\rightarrow
2 \, , \, 3 , \, 1
\;.
\label{01721}
\end{equation}
Therefore,
as in the case of the schemes A and B for four neutrinos,
the bounds on the LBL oscillation probabilities
are the same in the 3-neutrino schemes
\textrm{I} and \textrm{II}. In the following we will concentrate
on scheme I.

The bounds on the vacuum LBL
oscillation probabilities
$
P^{(\mathrm{LBL})}_{\stackrel{\makebox[0pt][l]
{$\hskip-3pt\scriptscriptstyle(-)$}}{\nu_{\alpha}}
\to\stackrel{\makebox[0pt][l]
{$\hskip-3pt\scriptscriptstyle(-)$}}{\nu_{\beta}}}
$
for the 4-neutrino schemes (\ref{AB})
are valid also in the case of mixing of three neutrinos:
the demonstrations
in the 4-neutrino case A (B)
can be carried over to the 3-neutrino case
\textrm{I} (\textrm{II})
if we put
$U_{\alpha4}=0$
($U_{\alpha1}=0$
and change the indices
$2,3,4\to1,2,3$)
for all
$\alpha=e,\mu,\tau$.
It is obvious that,
with
$ A_{\alpha;\beta} = 4 |U_{\beta 3}|^2 |U_{\alpha 3}|^2 $,
the same bounds on
$
P^{(\mathrm{LBL})}_{\stackrel{\makebox[0pt][l]
{$\hskip-3pt\scriptscriptstyle(-)$}}{\nu_{\alpha}}
\to\stackrel{\makebox[0pt][l]
{$\hskip-3pt\scriptscriptstyle(-)$}}{\nu_{\beta}}}
$
arise for $\alpha = \beta$ and $\alpha \neq \beta$
as given by Eqs.(\ref{paa})
and (\ref{pab1}).

It is interesting to observe that in the 3-neutrino case there
are no matter corrections,
apart from those of order
$a_{CC}/\Delta m^2$ which have been neglected
also in the 4-neutrino case.
This is easily understood by noting that
the matrix $R$ (\ref{block}) has now a $1 \times 1$ block
$R_{\mathrm{sun}}$. Consequently, there is no analogue to the
eigenvalue $\epsilon_4$ of $R$. This situation corresponds to
$ \omega = 0 $
and vanishing matter corrections at the order we are interested in.
Let us emphasize that this absence of matter corrections
is relative to the
bounds on the oscillation probabilities which we are discussing,
but in general
the oscillation probabilities
are affected by matter effects
through
$R_{\mathrm{atm}}$
and
$ \epsilon_2 - \epsilon_1 $.

In the following we will give the bounds on the LBL oscillation probabilities
for each of the
regions (\ref{regions}),
along the lines of the 4-neutrino section \ref{Vacuum Bounds}.

\emph{Region 1.}
With respect to SBL and LBL neutrino
oscillations,
the 3-neutrino schemes \textrm{I} and \textrm{II}
in region 1
correspond to the
4-neutrino schemes A and B, respectively,
with the same bounds on
$
P^{(\mathrm{LBL})}_{\stackrel{\makebox[0pt][l]
{$\hskip-3pt\scriptscriptstyle(-)$}}{\nu_{e}}
\to\stackrel{\makebox[0pt][l]
{$\hskip-3pt\scriptscriptstyle(-)$}}{\nu_{e}}}
$
(Eq.(\ref{081}) and Fig. \ref{fig1}) and
$
P^{(\mathrm{LBL})}_{\stackrel{\makebox[0pt][l]
{$\hskip-3pt\scriptscriptstyle(-)$}}{\nu_{\mu}}
\to\stackrel{\makebox[0pt][l]
{$\hskip-3pt\scriptscriptstyle(-)$}}{\nu_{e}}}
$
(Eqs.(\ref{pab12}) and (\ref{bbb1}) and Figs. \ref{fig2} and \ref{fig3}).

For completeness,
we want to mention that there is
a change
in the upper bound for
$
P^{(\mathrm{LBL})}_{\stackrel{\makebox[0pt][l]
{$\hskip-3pt\scriptscriptstyle(-)$}}{\nu_{\mu}}
\to\stackrel{\makebox[0pt][l]
{$\hskip-3pt\scriptscriptstyle(-)$}}{\nu_{e}}}
$
in going from four to three
neutrinos:
taking into account
the inequality $ c_e + c_\mu \geq 1 $,
we have
$ c_\mu \geq 1 - \min(a^0_e,a^0_\mu) $
and Eq.(\ref{bbb1}) improves to
\begin{equation}
P^{(\mathrm{LBL})}_{\stackrel{\makebox[0pt][l]
{$\hskip-3pt\scriptscriptstyle(-)$}}{\nu_{\mu}}
\to\stackrel{\makebox[0pt][l]
{$\hskip-3pt\scriptscriptstyle(-)$}}{\nu_{e}}}
\leq
a^0_e + (1-2a^0_e) \, \min(a^0_e,a^0_\mu)
\;.
\label{newbound1}
\end{equation}
For
$ a^0_e < a^0_\mu $
this bound is slightly more stringent than that given by
Eq.(\ref{0421}),
but the improvement is negligible for
$ a^0_e \ll 1 $.

\emph{Region 2.}
Actually, this region is excluded by the results of the
LSND experiment (see Refs. \cite{BBGK,BGKP,BGG96}) apart from
a small interval
of $\Delta m^2$ which might be marginally allowed.
The reason is that
(in combination with other data)
the upper bound
\begin{equation}\label{Ab}
A_{\mu;e} \leq 4\, a^0_e a^0_\mu
\end{equation}
is too restrictive to be compatible with the LSND data.
In spite of this evidence,
let us discuss the bounds
on the LBL probabilities in
this region.

The restrictions
$ c_e \geq 1-a^0_e $,
$ c_\mu \geq 1-a^0_\mu $
and the unitarity of the mixing matrix
imply that $ c_\tau $ is small:
$
c_\tau
=
2 - c_e - c_\mu
\leq
a^0_e + a^0_\mu
$.
From Eq.(\ref{pab1})
it follows that the probabilities of
$
\stackrel{\makebox[0pt][l]
{$\hskip-3pt\scriptscriptstyle(-)$}}{\nu_{\mu}}
\to
\stackrel{\makebox[0pt][l]
{$\hskip-3pt\scriptscriptstyle(-)$}}{\nu_{\tau}}
$
and
$
\stackrel{\makebox[0pt][l]
{$\hskip-3pt\scriptscriptstyle(-)$}}{\nu_{e}}
\to
\stackrel{\makebox[0pt][l]
{$\hskip-3pt\scriptscriptstyle(-)$}}{\nu_{\tau}}
$
transitions in LBL experiments
are confined in the range
\begin{equation}
\frac{1}{4} \, A_{\alpha;\tau}
\leq
P^{(\mathrm{LBL})}_{\stackrel{\makebox[0pt][l]
{$\hskip-3pt\scriptscriptstyle(-)$}}{\nu_{\alpha}}
\to\stackrel{\makebox[0pt][l]
{$\hskip-3pt\scriptscriptstyle(-)$}}{\nu_{\tau}}}
\leq
\frac{1}{4} \, A_{\alpha;\tau}
+ a^0_e + a^0_\mu
\qquad ( \alpha = e , \mu )
\;,
\end{equation}
whereas for the probability of
$
\stackrel{\makebox[0pt][l]
{$\hskip-3pt\scriptscriptstyle(-)$}}{\nu_{\mu}}
\to\stackrel{\makebox[0pt][l]
{$\hskip-3pt\scriptscriptstyle(-)$}}{\nu_{e}}
$
transitions we have only the lower bound
\begin{equation}
\frac{1}{4} \, A_{\mu;e}
\leq
P^{(\mathrm{LBL})}_{\stackrel{\makebox[0pt][l]
{$\hskip-3pt\scriptscriptstyle(-)$}}{\nu_{\mu}}
\to\stackrel{\makebox[0pt][l]
{$\hskip-3pt\scriptscriptstyle(-)$}}{\nu_{e}}}
\;.
\end{equation}
The inequality (\ref{pab3}) leads to the
additional upper bounds
\begin{equation}
P^{(\mathrm{LBL})}_{\stackrel{\makebox[0pt][l]
{$\hskip-3pt\scriptscriptstyle(-)$}}{\nu_{\alpha}}
\to\stackrel{\makebox[0pt][l]
{$\hskip-3pt\scriptscriptstyle(-)$}}{\nu_{\tau}}}
\leq
\left( a^0_e + a^0_\mu \right)
\left( 2 - a^0_e - a^0_\mu \right)
\qquad ( \alpha = e , \mu )
\;.
\end{equation}

We want to mention that the scenario of Ref. \cite{AP97} is
settled in region 2 and seems to take advantage of the fact that
$ \Delta m^2 \simeq 1.7 \, \mathrm{eV}^2 $
is marginally allowed despite of
Eq.(\ref{Ab}) (see Refs. \cite{BGG96,baksan}).
In this way Ref. \cite{AP97}
incorporates the LSND data whereas the atmospheric
neutrino anomaly and the solar neutrino deficit are taken into
account by a single
$ \Delta m^2_{\mathrm{atm, sun}} \sim 10^{-2} \, \mathrm{eV}^2 $,
which
allows for the zenith angle variation of the atmospheric
neutrino anomaly
but leads to an energy-independent suppression of the
solar neutrino flux that is unfavoured
by the data of
solar neutrino experiments \cite{petcov,CMMV97}.
Moreover,
it has been shown in Ref. \cite{FLMS97}
that a combined analysis of all SBL data
excludes this scenario at $\sim 99\%$ CL.

\emph{Region 3.}
In this region,
where $c_e \geq 1-a^0_e$ and
$c_\mu \leq a^0_\mu$,
the full set of atmospheric neutrino data cannot be explained
in the framework discussed here.
The reason is that for sub-GeV events one has
\begin{equation}
P^{(\mathrm{LBL})}_{\stackrel{\makebox[0pt][l]
{$\hskip-3pt\scriptscriptstyle(-)$}}{\nu_{\mu}}
\to\stackrel{\makebox[0pt][l]
{$\hskip-3pt\scriptscriptstyle(-)$}}{\nu_{\mu}}}
\geq (1-a^0_\mu)^2
\end{equation}
and this is incompatible \cite{BGG96}
with the atmospheric neutrino anomaly except for values of
$\Delta m^2$ close to 0.3 eV$^2$,
where there is no zenith-angle variation.
The LBL transition probabilities 
of muon neutrinos are confined by
\begin{equation}
\frac{1}{4} \, A_{\mu;\beta}
\leq
P^{(\mathrm{LBL})}_{\stackrel{\makebox[0pt][l]
{$\hskip-3pt\scriptscriptstyle(-)$}}{\nu_{\mu}}
\to\stackrel{\makebox[0pt][l]
{$\hskip-3pt\scriptscriptstyle(-)$}}{\nu_{\beta}}}
\leq
\frac{1}{4} \, A_{\mu;\beta} + a^0_\mu
\qquad
( \beta = e , \tau )
\;,
\end{equation}
whereas for
$
\stackrel{\makebox[0pt][l]
{$\hskip-3pt\scriptscriptstyle(-)$}}{\nu_{e}}
\to\stackrel{\makebox[0pt][l]
{$\hskip-3pt\scriptscriptstyle(-)$}}{\nu_{\tau}}
$
transitions
there is only the
lower bound
\begin{equation}
\frac{1}{4} \, A_{e;\tau}
\leq
P^{(\mathrm{LBL})}_{\stackrel{\makebox[0pt][l]
{$\hskip-3pt\scriptscriptstyle(-)$}}{\nu_{e}}
\to\stackrel{\makebox[0pt][l]
{$\hskip-3pt\scriptscriptstyle(-)$}}{\nu_{\tau}}}
\;.
\end{equation}
The inequality (\ref{pab2}),
which is a consequence of
probability conservation,
leads to
\begin{equation}
P^{(\mathrm{LBL})}_{\stackrel{\makebox[0pt][l]
{$\hskip-3pt\scriptscriptstyle(-)$}}{\nu_{\mu}}
\to\stackrel{\makebox[0pt][l]
{$\hskip-3pt\scriptscriptstyle(-)$}}{\nu_{\beta}}}
\leq
a^0_\mu \left( 2 - a^0_\mu \right) 
\qquad
( \beta = e , \tau )
\;.
\label{oldbound3}
\end{equation}
Furthermore,
taking into account
the inequality $ c_e + c_\mu \geq 1 $,
we have
$ c_e \geq 1 - \min(a^0_e,a^0_\mu) $
and Eq.(\ref{bbb1}) improves to
\begin{equation}
P^{(\mathrm{LBL})}_{\stackrel{\makebox[0pt][l]
{$\hskip-3pt\scriptscriptstyle(-)$}}{\nu_{\mu}}
\to\stackrel{\makebox[0pt][l]
{$\hskip-3pt\scriptscriptstyle(-)$}}{\nu_{e}}}
\leq
a^0_\mu + (1-2a^0_\mu) \, \min(a^0_e,a^0_\mu)
\;.
\label{newbound3}
\end{equation}
For
$ a^0_e \ll a^0_\mu \ll 1 $
this bound is about half of that given by
Eq.(\ref{oldbound3}).

The 3-neutrino scheme of Ref. \cite{CF97-MR97},
which lies in region 3,
merges the LSND and atmospheric mass-squared scales and
dispenses with the zenith angle variation of the atmospheric
neutrino anomaly. This is only allowed at
$ \Delta m^2 \simeq 0.3 \, \mathrm{eV}^2 $
(see also Ref. \cite{FLMS97}).
If one accepts this possibility,
the low mass-squared
difference is free to be used for a solution of the solar
neutrino deficit problem.

The differences in the bounds on the LBL
probabilities are marked and could thus serve
to distinguish between the three
different regions in the 3-neutrino case.
They also serve as a cross-check
for present hints of neutrino oscillations.
Of course, in the
experiments discussed the bounds on the transition probabilities in vacuum
in the 4-neutrino case (schemes A and B)
are indistinguishable from those in the
3-neutrino case with region 1.
However,
LBL experiments could distinguish the 4-neutrino case
from the 3-neutrino case
by measuring a transition probability
$
P^{(\mathrm{SBL})}_{\stackrel{\makebox[0pt][l]
{$\hskip-3pt\scriptscriptstyle(-)$}}{\nu_{\mu}}
\to\stackrel{\makebox[0pt][l]
{$\hskip-3pt\scriptscriptstyle(-)$}}{\nu_{e}}}
$
or a survival probability
$P^{(\mathrm{LBL})}_{\bar\nu_{e}\to\bar\nu_{e}}$
which is incompatible with the vacuum bounds
but satisfies the bounds in matter obtained
in the 4-neutrino case.
Such an observation would exclude
the
3-neutrino case with region 1.

\section{Conclusions}
\label{Conclusions}

At present there are three experimental indications
in favour of neutrino
oscillations which correspond to three different scales of
neutrino mass-squared differences:
the solar neutrino deficit, the atmospheric neutrino
anomaly and the result of the LSND experiment.
These indications
and the negative results of numerous short-baseline neutrino
experiments can be accommodated in two
schemes (A and B) with mixing of four massive neutrinos
\cite{BGG96}.
In this paper
we have presented a detailed study of the
predictions of the schemes A and B for
long-baseline experiments.
We have discussed what general conclusions
on the long-baseline transition probabilities
between different neutrino states
can be inferred from the existing data 
of short-baseline experiments,
taking into account
the results of solar and atmospheric neutrino experiments.
We have obtained rather strong bounds on the
probabilities of 
$ \bar\nu_e \to \bar\nu_e $
and
$
\stackrel{\makebox[0pt][l]
{$\hskip-3pt\scriptscriptstyle(-)$}}{\nu_{\mu}}
\to\stackrel{\makebox[0pt][l]
{$\hskip-3pt\scriptscriptstyle(-)$}}{\nu_{e}}
$
LBL transitions.
Matter effects were thoroughly taken into account and we have
shown that they do not change substantially the main conclusions
drawn from the vacuum case.

The schemes A and B give completely different predictions for
neutrinoless double beta decay and for neutrino mass effects in
experiments for neutrino mass measurements by the tritium method
\cite{BGG96}.
They lead, however,
to the same bounds on long-baseline oscillation probabilities.
In addition, all the bounds that we have derived
apply for neutrinos as well as antineutrinos.

We have shown that the results of the
short-baseline reactor experiments put
rather severe bounds on the probability
$1-P^{(\mathrm{LBL})}_{\bar\nu_e \to \bar\nu_e}$
of $\bar\nu_e$ transitions
into all possible other states
in the long-baseline CHOOZ and Palo Verde reactor experiments.
If the $\Delta{m}^2$ relevant in short-baseline oscillations
is bigger than about
$ 3 \, \mathrm{eV}^2 $,
the bound on
$1-P^{(\mathrm{LBL})}_{\bar\nu_e \to \bar\nu_e}$
is slightly higher
than the sensitivity of the CHOOZ experiment,
allowing some possibility to
reveal neutrino oscillations in this channel.
However,
the results of the
LSND experiment favour the range
$ 0.3 \lesssim \Delta{m}^2 \lesssim 2.2 \, \mathrm{eV}^2 $.
We have
shown that in this range the
upper bound for the quantity 
$1-P^{(\mathrm{LBL})}_{\bar\nu_e\to\bar\nu_e}$
lies between
$10^{-2}$ and $5\times10^{-2}$
(see Fig. \ref{fig1}) and thus is below the sensitivity
of CHOOZ and Palo Verde.

The Kam-Land reactor experiment is very interesting because
its baseline of 150 km is very long compared to the baseline
of 1 km of CHOOZ and
Palo Verde.
If the solar neutrino deficit
problem is to be resolved by vacuum oscillations,
the situation
is very similar to the other LBL reactor experiments.
The upper
bound on
$1-P^{(\mathrm{LBL})}_{\bar\nu_e\to\bar\nu_e}$
is shown in
Fig. \ref{fig4} by the solid line.
It deviates from the short-dashed curve,
which is valid in vacuum,
because matter corrections are not negligible though small in this case,
due to the long baseline.
The bound represented by the short-dashed curve
in Fig. \ref{fig4}
is valid also in the case of a small mixing angle MSW
solution of the solar neutrino problem.
On the other hand,
we have shown in section
\ref{sim}
that,
if large mixing angle MSW resonant
flavour transitions are responsible for the solar
neutrino deficit,
the bound on
$1-P^{(\mathrm{LBL})}_{\bar\nu_e\to\bar\nu_e}$
in Kam-Land
becomes practically trivial
at small neutrino energies
($\sim1\,\mathrm{MeV}$).
Therefore, if in the
Kam-Land experiment a large value of
$1-P^{(\mathrm{LBL})}_{\bar\nu_e\to\bar\nu_e}$
is found at small neutrino energies,
with a suppression at higher neutrino energies,
we would have an indirect indication in favour of
the large mixing angle
MSW solution of the solar neutrino deficit.

%
We have also shown that in this case
$\Delta{m}_{\mathrm{sun}}^2$ could be
determined by measuring the maxima of 
$P^{(\mathrm{LBL})}_{\bar\nu_e\to\bar\nu_e}$ as a function of the
neutrino energy. There are two conditions that
must be satisfied in a LBL experiment
in order to have the possibility to measure
$\Delta m^2_{\mathrm{sun}}$, apart from the necessary
sensitivity and energy resolution. First one needs 
$\Delta m^2_{\mathrm{sun}} \gg a_{CC}$ in
order that this measurement is not disturbed by matter
effects. Second, $\Delta m^2_{\mathrm{sun}}L/2p \sim 1$ is
required. These conditions lead to $p \ll 40$ MeV and 
$L \sim 40 \times (p/1 \, \mathrm{MeV})$ km for $\Delta m^2_{\mathrm{sun}}
\sim 10^{-5}$ eV$^2$. At present, among the planned LBL experiments 
these conditions are only met by Kam-Land.

In Figs. \ref{fig2} and \ref{fig3} the solid lines depict the
upper bounds on the probability
$
P^{(\mathrm{LBL})}_{\stackrel{\makebox[0pt][l]
{$\hskip-3pt\scriptscriptstyle(-)$}}{\nu_{\mu}}
\to\stackrel{\makebox[0pt][l]
{$\hskip-3pt\scriptscriptstyle(-)$}}{\nu_{e}}}
$
of
$
\stackrel{\makebox[0pt][l]
{$\hskip-3pt\scriptscriptstyle(-)$}}{\nu_{\mu}}
\to\stackrel{\makebox[0pt][l]
{$\hskip-3pt\scriptscriptstyle(-)$}}{\nu_{e}}
$
transitions for the K2K experiment and the MINOS and ICARUS
experiments, respectively.
In the derivation of the solid curves
matter effects are included and thus
in the schemes A and B such transitions are severely constrained
by the results of short-baseline
reactor and accelerator experiments.
The sensitivities of MINOS and ICARUS
is well below the upper bound for
$
P^{(\mathrm{LBL})}_{\stackrel{\makebox[0pt][l]
{$\hskip-3pt\scriptscriptstyle(-)$}}{\nu_{\mu}}
\to\stackrel{\makebox[0pt][l]
{$\hskip-3pt\scriptscriptstyle(-)$}}{\nu_{e}}}
$
whereas the sensitivity of
the K2K experiment might be insufficient.
If matter effects are neglected,
the upper bound on this
probability is given by the short-dashed lines in
Figs. \ref{fig2} and \ref{fig3}.
In all four figures the shadowed
regions (light and dark) indicate the range (\ref{range})
determined by the LSND experiment and the negative results of
all the other SBL experiments.

We have shown that there is also an
upper bound on long-baseline
$
\stackrel{\makebox[0pt][l]
{$\hskip-3pt\scriptscriptstyle(-)$}}{\nu_{e}}
\to\stackrel{\makebox[0pt][l]
{$\hskip-3pt\scriptscriptstyle(-)$}}{\nu_{\tau}}
$
oscillations which is less tight than the one 
for
$
\stackrel{\makebox[0pt][l]
{$\hskip-3pt\scriptscriptstyle(-)$}}{\nu_{\mu}}
\to\stackrel{\makebox[0pt][l]
{$\hskip-3pt\scriptscriptstyle(-)$}}{\nu_{e}}
$
transitions. It is indicated by the long-dashed curves in
Figs. \ref{fig2} and \ref{fig3} which would be relevant if the corresponding
experiments would use a 
$\stackrel{\makebox[0pt][l]
{$\hskip-3pt\scriptscriptstyle(-)$}}{\nu_{e}}$ beam.
On the other hand,
the long-baseline
$
\stackrel{\makebox[0pt][l]
{$\hskip-3pt\scriptscriptstyle(-)$}}{\nu_{\mu}}
\to\stackrel{\makebox[0pt][l]
{$\hskip-3pt\scriptscriptstyle(-)$}}{\nu_{\mu}}
$
and
$
\stackrel{\makebox[0pt][l]
{$\hskip-3pt\scriptscriptstyle(-)$}}{\nu_{\mu}}
\to\stackrel{\makebox[0pt][l]
{$\hskip-3pt\scriptscriptstyle(-)$}}{\nu_{\tau}}
$
channels are unconstrained with the methods discussed here.

We have obtained bounds on LBL transition probabilities
in the case of the neutrino mass spectra (\ref{AB}),
which are implied by the results
of the solar, atmospheric and
LSND experiments.
If the LSND data are not confirmed by future experiments,
but nevertheless there is a mass (or masses) approximately
equal to 1 eV
providing an explanation for
the hot dark matter problem,
then the
neutrino mass spectrum can be different from
the spectra A and B in
Eq.(\ref{AB}).
The natural neutrino mass spectrum in this case is hierarchical
and the bounds
that we have obtained in this
paper are not valid.

We have also made a digression to 3-neutrino scenarios and
discussed the bounds on the transition probabilities for all possible
cases such that the two mass-squared differences correspond to
SBL and LBL neutrino oscillations.
We have argued that for three
neutrinos matter corrections
to the bounds on the transition probabilities
are absent,
apart from those of order
$a_{CC}/\Delta m^2$ which have been neglected
also in the 4-neutrino case.

Summarizing,
we would like to emphasize that 
the results of all neutrino oscillation experiments
lead to severe
constraints for the
probabilities of
$\bar\nu_e$
disappearance and
$
\stackrel{\makebox[0pt][l]
{$\hskip-3pt\scriptscriptstyle(-)$}}{\nu_{\mu}}
\to\stackrel{\makebox[0pt][l]
{$\hskip-3pt\scriptscriptstyle(-)$}}{\nu_{e}}
$
appearance
in long-baseline experiments.
Nevertheless, the allowed region for the probability in the
$
\stackrel{\makebox[0pt][l]
{$\hskip-3pt\scriptscriptstyle(-)$}}{\nu_{\mu}}
\to\stackrel{\makebox[0pt][l]
{$\hskip-3pt\scriptscriptstyle(-)$}}{\nu_{e}}
$
channel
is well within the planned
sensitivities of the MINOS and ICARUS experiments.
The channels
$
\stackrel{\makebox[0pt][l]
{$\hskip-3pt\scriptscriptstyle(-)$}}{\nu_{\mu}}
\to\stackrel{\makebox[0pt][l]
{$\hskip-3pt\scriptscriptstyle(-)$}}{\nu_{\tau}}
$
and
$
\stackrel{\makebox[0pt][l]
{$\hskip-3pt\scriptscriptstyle(-)$}}{\nu_{\mu}}
\to\stackrel{\makebox[0pt][l]
{$\hskip-3pt\scriptscriptstyle(-)$}}{\nu_{\mu}}
$
are not constrained at all.
Therefore,
from the point of view of the present investigation, 
long-baseline muon neutrino beams provide promising
facilities for the observation of neutrino oscillations.
However,
it is important to note that
future measurements by LBL experiments of
$ \bar\nu_e \to \bar\nu_e $
and/or
$
\stackrel{\makebox[0pt][l]
{$\hskip-3pt\scriptscriptstyle(-)$}}{\nu_{\mu}}
\to\stackrel{\makebox[0pt][l]
{$\hskip-3pt\scriptscriptstyle(-)$}}{\nu_{e}}
$
transition probabilities
that violate the bounds presented in this paper
would allow to exclude the 4-neutrino schemes (\ref{AB}).

\newpage

\begin{center}
\textbf{NOTE ADDED}
\end{center}

After
we finished this paper the results of the first
long-baseline reactor experiment CHOOZ appeared
(M. Apollonio \emph{et al.}, preprint hep-ex/9711002).
No indications in favor of
$\bar\nu_e\to\bar\nu_e$
transitions were found in this experiment.
The upper bound for the
transition probability
of electron antineutrinos into other states
found in the CHOOZ experiment is in agreement with
the limit presented in Fig.\ref{fig1}.

\acknowledgments

C.G. would like to thank
K. Inoue,
E. Lisi,
F. Martelli,
H. Nunokawa,
O. Peres,
A. Rossi,
V. Semikoz
and
F. Vetrano
for useful discussions at TAUP97.
S.M.B. like to acknowledge support from Dyson Visiting
Professor Funds at the Institute for Advanced Study.

\begin{table}
\begin{tabular}{lcccc} 
Experiment & $\langle p \rangle/ 1 \: \mathrm{GeV}$ & $L/1 \:
\mathrm{km}$ & $a_{CC}/1 \: \mathrm{eV}^2$ & $\omega_{\mathrm{max}}$
\\ \hline
Kam-Land (vac. osc.) & $10^{-3}$ & 150 & $2.3 \times 10^{-7}$ & $0.13$ \\
K2K & 1 & 250 & $2.3 \times 10^{-4}$ & 0.22 \\
MINOS & 10 & 730 & $2.3 \times 10^{-3}$ & 0.63 \\
ICARUS & 25 & 730 & $5.8 \times 10^{-3}$ & 0.63 
\end{tabular}
\caption{\label{tab1}
List of the 
planned LBL experiments (except CHOOZ and Palo Verde where
matter effects are absent) with their
average neutrino momenta $\langle p \rangle$, the length $L$ of the
baseline, the value of the matter parameter $a_{CC}$ and
the maximal value of the phase $\omega$
(given by Eq.(\ref{omax}))
characterizing the matter effects in
the bounds on LBL neutrino oscillation probabilities.}
\end{table}

\begin{figure}[h]
\refstepcounter{figure}
\label{fig1}
FIG.\ref{fig1}.
Upper bound for
the transition probability
$1-P^{(\mathrm{LBL})}_{\bar\nu_e\to\bar\nu_e}$
in the CHOOZ and Palo Verde experiments (solid curve),
for
$\Delta{m}^2$
in the range
$
10^{-1} \, \mathrm{eV}^2
\leq \Delta{m}^2 \leq
10^{3} \, \mathrm{eV}^2
$.
The upper bound was obtained 
from the 90\% CL exclusion plot of the Bugey
$\bar\nu_e\to\bar\nu_e$ experiment \cite{Bugey95}.
The dash-dotted and dash-dot-dotted vertical lines
depict, respectively, the expected sensitivities
of the CHOOZ and Palo Verde
LBL reactor neutrino experiments.
The shadowed region
corresponds to the range of $\Delta{m}^2$
allowed at 90\% CL by the results of the LSND experiment,
taking into account the results of
all the other SBL experiments
(see Eq.(\ref{range})).
\end{figure}

\begin{figure}[h]
\refstepcounter{figure}
\label{fig2}
FIG.\ref{fig2}.
Upper bounds for
the probability 
of
$\nu_{\mu}\to\nu_{e}$ 
transitions in the K2K experiment.
The solid curve is obtained by a numerical analysis of
Eq.(\ref{omeb}) and uses the following experimental input:
the 90\% CL exclusion plot of the Bugey
$\bar\nu_e\to\bar\nu_e$ experiment \cite{Bugey95},
the 90\% CL exclusion plots of the
BNL E734 \cite{BNLE734},
BNL E776 \cite{BNLE776} and
CCFR \cite{CCFR97}
$
\stackrel{\makebox[0pt][l]
{$\hskip-3pt\scriptscriptstyle(-)$}}{\nu_{\mu}}
\to\stackrel{\makebox[0pt][l]
{$\hskip-3pt\scriptscriptstyle(-)$}}{\nu_{e}}
$
experiments
and
the 90\% CL exclusion plots
of the CDHS \cite{CDHS84} and CCFR \cite{CCFR84}
$\nu_{\mu}\to\nu_{\mu}$
experiments.
The solid curve is the matter-corrected version of
the short-dashed curve,
which represents the bound (\ref{pab12})
valid for neutrino oscillations in vacuum
(this curve does not need
the input of the
$\nu_{\mu}\to\nu_{\mu}$
experiments).
The long-dashed line represents the bound (\ref{abba})
derived from probability conservation
and has been
evaluated by using the $\bar\nu_e\to\bar\nu_e$
data.
The dotted curve depicts the ``matter-stable'' bound
(\ref{CSee}),
which needs experimental input from
$\bar\nu_e\to\bar\nu_e$
and
$\nu_{\mu}\to\nu_{\mu}$
transitions.
The dash-dotted vertical line represents
the expected sensitivity 
of the LBL accelerator neutrino experiment K2K.
The shadowed region
corresponds to the range of mixing parameters
allowed at 90\% CL by the results of the LSND experiment,
taking into account the results of
all the other SBL experiments.
The two horizontal borderlines correspond
to the limits (\ref{range}) for $\Delta m^2$.
The darkly shadowed area represents the allowed region if matter
effects are neglected. The left edge of this region is given by
the lower bound Eq.(\ref{pmin}) of LSND on the probability of 
$
\stackrel{\makebox[0pt][l]
{$\hskip-3pt\scriptscriptstyle(-)$}}{\nu_{\mu}}
\to\stackrel{\makebox[0pt][l]
{$\hskip-3pt\scriptscriptstyle(-)$}}{\nu_{e}}
$ transitions.
The long-dashed curve constitutes also an upper bound for
the probability
of
$
\stackrel{\makebox[0pt][l]
{$\hskip-3pt\scriptscriptstyle(-)$}}{\nu_{e}}
\to\stackrel{\makebox[0pt][l]
{$\hskip-3pt\scriptscriptstyle(-)$}}{\nu_{\tau}}
$
transitions if K2K would use a
$
\stackrel{\makebox[0pt][l]
{$\hskip-3pt\scriptscriptstyle(-)$}}{\nu_{e}}
$
beam.
\end{figure}

\begin{figure}[h]
\refstepcounter{figure}
\label{fig3}
FIG.\ref{fig3}.
The same as in Fig. \ref{fig2} but the matter corrections now
refer to the MINOS and ICARUS experiments with the dot-dashed
and dot-dot-dashed lines as their respective sensitivities.
\end{figure}

\begin{figure}[h]
\refstepcounter{figure}
\label{fig4}
FIG.\ref{fig4}.
Upper bounds for
the transition probability
$1-P^{(\mathrm{LBL})}_{\bar\nu_e\to\bar\nu_e}$
in the Kam-Land experiment.
The short dashed curve (see Eq.(\ref{081}))
represents the bound for vacuum oscillations
(it is identical with the solid curve in Fig. \ref{fig1})
and is valid also in matter
if the solar neutrino problem
is explained by
the small mixing angle MSW solution.
The solid curve 
represents the bound (\ref{eb}) valid in matter
with the value of $\omega_{\mathrm{max}}$
given in Tab. \ref{tab1} and it refers
to the case of a vacuum oscillation solution
of the solar neutrino problem.
The dotted, dash-dotted and dash-dot-dotted lines
give the upper bounds for
$1-P^{(\mathrm{LBL})}_{\bar\nu_e\to\bar\nu_e}$
at different neutrino momenta $p$
in the case of a large mixing angle solution of
the solar neutrino problem
(see Eq.(\ref{lmsw})).
The long-dashed vertical line
depicts the expected sensitivity
of the Kam-Land experiment.
The shadowed region and the two horizontal solid lines
correspond to the range of $\Delta{m}^2$
allowed at 90\% CL by the results of the LSND experiment,
taking into account the results of
all the other SBL experiments
(see Eq.(\ref{range})).
\end{figure}


\newpage

\begin{minipage}[p]{0.95\textwidth}
\begin{center}
\mbox{\epsfig{file=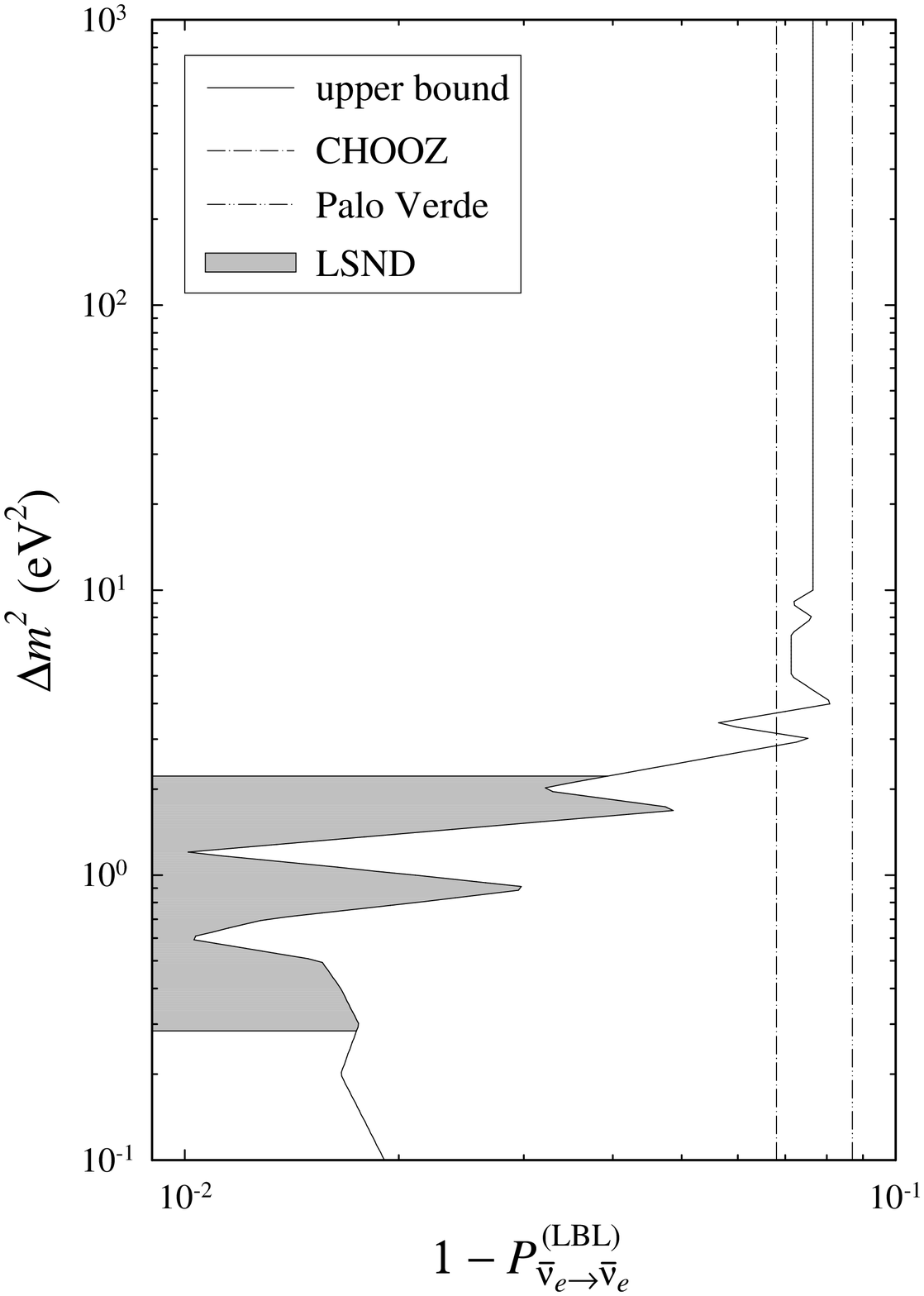,height=0.95\textheight}}
\end{center}
\end{minipage}
\begin{center}
\Large Figure~\ref{fig1}
\end{center}

\newpage

\begin{minipage}[p]{0.95\textwidth}
\begin{center}
\mbox{\epsfig{file=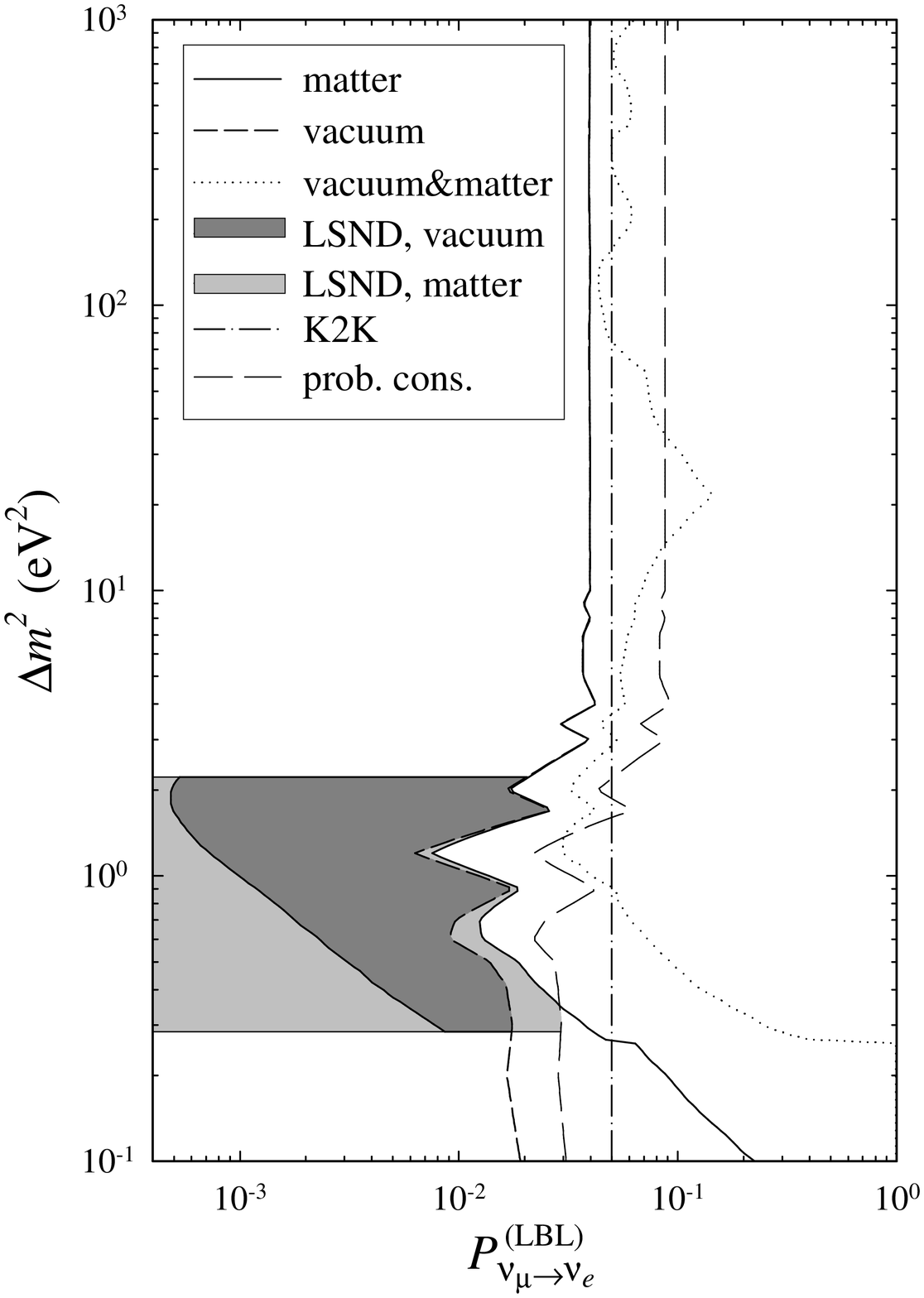,height=0.95\textheight}}
\end{center}
\end{minipage}
\begin{center}
\Large Figure~\ref{fig2}
\end{center}

\newpage

\begin{minipage}[p]{0.95\textwidth}
\begin{center}
\mbox{\epsfig{file=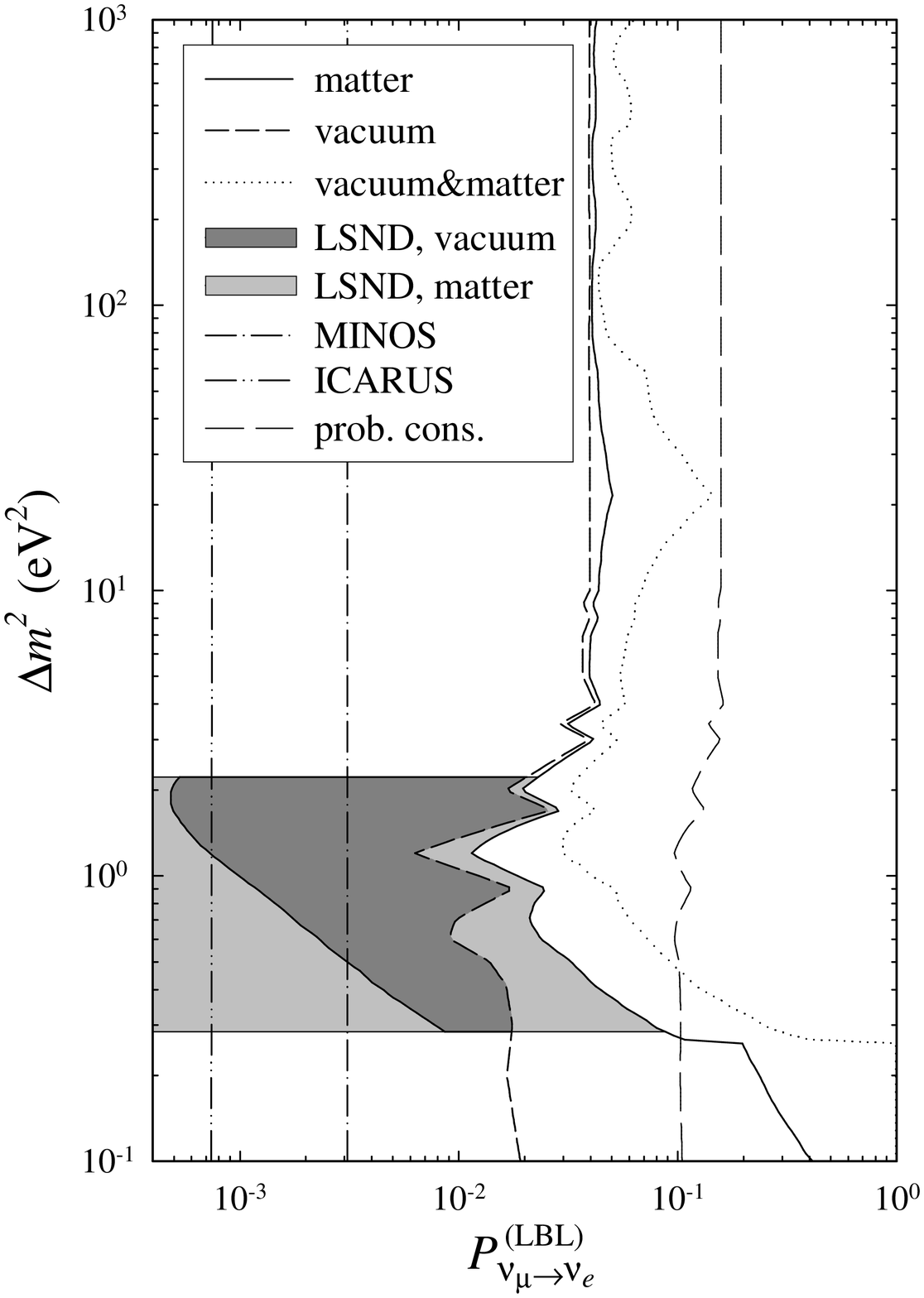,height=0.95\textheight}}
\end{center}
\end{minipage}
\begin{center}
\Large Figure~\ref{fig3}
\end{center}

\newpage

\begin{minipage}[p]{0.95\textwidth}
\begin{center}
\mbox{\epsfig{file=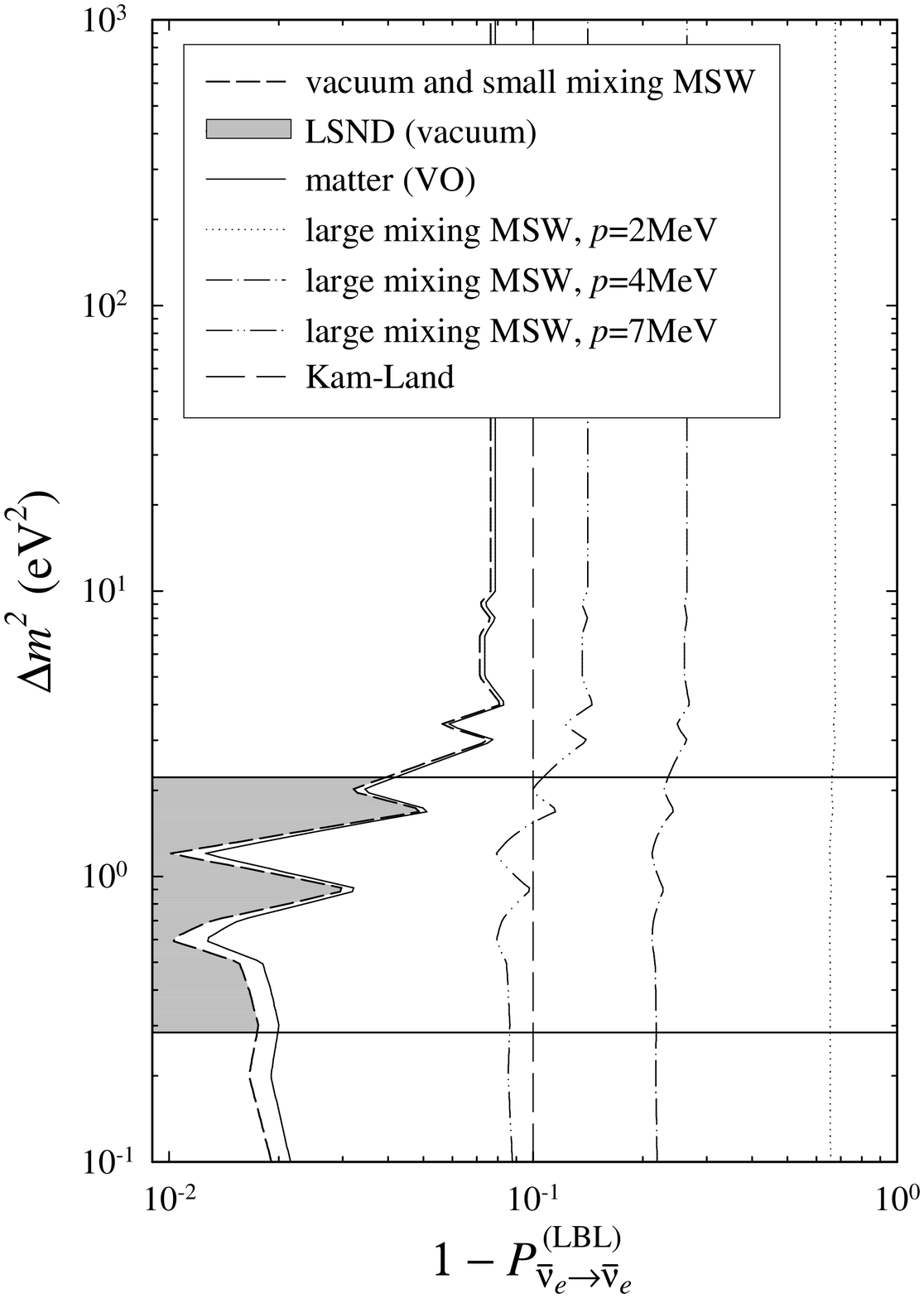,height=0.95\textheight}}
\end{center}
\end{minipage}
\begin{center}
\Large Figure~\ref{fig4}
\end{center}

\end{document}